\documentclass[a4paper,twocolumn,11pt,accepted=2024-02-24]{quantumarticle}
\pdfoutput=1
\usepackage[numbers]{natbib}

\usepackage{graphicx}
\usepackage{dcolumn}
\usepackage{bm}
\usepackage{algorithm}
\usepackage{algpseudocode}
\usepackage{float}
\usepackage{braket}
\usepackage{lipsum}
\usepackage{orcidlink}
\usepackage[hashEnumerators,smartEllipses]{markdown}

\usepackage[caption=false,listofformat=subsimple,labelformat=simple]{subfig}

\usepackage[export]{adjustbox}

\usepackage{textcomp} 

\usepackage{tabu}

\usepackage{color}
\definecolor{darkred}{RGB}{150,0,0}

\usepackage{xparse}
\RenewDocumentCommand\log{g}{%
  \IfNoValueF{#1}{\text{log}\left(#1\right)}%
  \IfValueF{#1}{\text{log}}%
}
\usepackage{hyperref}
\newcommand{\etal}{\textit{et al.} }

\newcommand{\edit}[1]{#1}

\DeclareMathOperator{\Tr}{Tr}

\begin{document}


\title{%
Quantum state preparation via engineered ancilla resetting
}%

\author{Daniel Alcalde Puente \orcidlink{0000-0002-3519-5931}} 
\affiliation{Forschungszentrum Jülich, Institute of Quantum Control, Peter Grünberg Institut (PGI-8), 52425 Jülich, Germany}
\affiliation{Institute for Theoretical Physics, University of Cologne, 50937 Köln, Germany}

\author{Felix Motzoi \orcidlink{0000-0003-4756-5976}}
\affiliation{Forschungszentrum Jülich, Institute of Quantum Control,
Peter Grünberg Institut (PGI-8), 52425 Jülich, Germany}

\author{Tommaso Calarco \orcidlink{0000-0001-5364-7316}}
\affiliation{Forschungszentrum Jülich, Institute of Quantum Control,
Peter Grünberg Institut (PGI-8), 52425 Jülich, Germany}
\affiliation{Institute for Theoretical Physics, University of Cologne, 50937 Köln, Germany}
\affiliation{Dipartimento di Fisica e Astronomia, Universit\'{a} di Bologna, 40127 Bologna, Italy}

\author{Giovanna Morigi \orcidlink{0000-0002-1946-3684}}
\affiliation{Theoretical Physics, Department of Physics, Saarland University, 66123 Saarbrücken, Germany}

\author{Matteo Rizzi \orcidlink{0000-0002-8283-1005}}
\affiliation{Forschungszentrum Jülich, Institute of Quantum Control,
Peter Grünberg Institut (PGI-8), 52425 Jülich, Germany}
\affiliation{Institute for Theoretical Physics, University of Cologne, 50937 Köln, Germany}


\begin{abstract}
In this theoretical investigation, we examine the effectiveness of a protocol incorporating periodic quantum resetting for preparing ground states of frustration-free parent Hamiltonians. This protocol uses a steering Hamiltonian that enables local coupling between the system and ancillary degrees of freedom. At periodic intervals, the ancillary system is reset to its initial state. For infinitesimally short reset times, the dynamics can be approximated by a Lindbladian whose steady state is the target state. For finite reset times, however, the spin chain and the ancilla become entangled between reset operations. 
To evaluate the protocol, we employ Matrix Product State simulations and quantum trajectory techniques, focusing on the preparation of the spin-1 Affleck-Kennedy-Lieb-Tasaki state. Our analysis considers convergence time, fidelity, and energy evolution under different reset intervals. Our numerical results show that ancilla system entanglement is essential for faster convergence. In particular, there exists an optimal reset time at which the protocol performs best. Using a simple approximation, we provide insights into how to optimally choose the mapping operators applied to the system during the reset procedure. Furthermore, the protocol shows remarkable resilience to small deviations in reset time and dephasing noise. Our study suggests that stroboscopic maps using quantum resetting may offer advantages over alternative methods, such as quantum reservoir engineering and quantum state steering protocols, which rely on Markovian dynamics.
\end{abstract}

\maketitle

\section{Introduction}
Quantum technologies hold immense potential for addressing significant challenges in quantum simulation, communication, and information processing. Consequently, the preparation of fiducial quantum states and the development of noise-resistant state preparation routines are of paramount importance, especially for noisy intermediate-scale quantum (NISQ) devices \cite{Preskill2018quantumcomputingin}.

Quantum state preparation by means of unitary quantum circuits often faces challenges due to the required circuit complexity \cite{eisert2021entangling}, which may demand large circuit depths and thus require low error rates, unattainable in the NISQ era. Various approaches have been proposed, starting from adiabatic techniques \cite{RevModPhys.90.015002, sompet2022realizing, wei2022efficient} and sequences of unitary transformations \cite{PhysRevLett.95.110503,motzoi2017linear}.

An alternative, and possibly more robust, strategy implements protocols based on non-unitary dynamics. One strategy goes under the name of reservoir engineering \cite{Poyatos:1996,Pielawa:2007,Diehl:2008,Verstrate:2009, schirmer2010stabilizing} and aims at designing effective master equations whose steady state is the target state of the quantum state preparation. Quantum reservoir engineering has been proposed for quantum computing \cite{Verstrate:2009} and for robust preparation of many-body quantum states \cite{Morigi:2015,zhou2021symmetry}.
{Designing Markovian master equations, however, limits to performing operations in the weak coupling limit, leading to slow convergence speeds, which could be counterbalanced by including time-dependent controls \cite{motzoi2016backaction}.

Further strategies pursue quantum state preparation using adaptive measurements, as for example in Ref.~\cite{smith2022deterministic, tantivasadakarn2021long}. These protocols are efficient, yet require the use of feedback based on non-local measurements, which is not yet widely accessible on current quantum platforms. In particular, local feedback control has only been achieved experimentally in a few cases, and it underperforms coherent entanglement gates by about two orders of magnitude \cite{sayrin2011real,vijay2012stabilizing,riste2013deterministic}.
 Preparing non-local states between spatially separated elements \cite{mabuchi2009continuous, kerckhoff2010designing, martin2015deterministic} or applying feedback to larger lattices \cite{google2023suppressing} is expected to create an even larger gap.

A procedure for preparing quantum states from a fiducial one, which is not adaptive and is based solely on local operations, can be achieved through quantum state steering via measurements and reset of ancillary degrees of freedom \cite{BurgarthPRA2007,BurgarthPRL2007,
matthies2022programmable,roy2020measurement}. It is known that for unitary dynamics the use of ancillas can dramatically improve the speed of unitary quantum circuits \cite{motzoi2017linear,moore2001parallel, van2005fast, draper2004logarithmic}. Likewise, coherent feedback is known to have improved best-case scaling compared to measurement-based approaches for state preparation \cite{jacobs2014coherent}. Here, we make use of ancilla coherence to accelerate dissipative dynamics, where the ancillas act as a non-Markovian memory within the engineered reservoir \cite{rivas2010entanglement}. 
Thus, in this study, we explore the transition from quantum reservoir engineering to quantum state steering via the central tool of a stroboscopic map, which we turn from representing just an unraveling of the Lindblad master equation to becoming the key to go beyond.

\begin{figure*}[t!]
    \centering
    \subfloat[]{\includegraphics[height=0.3\textwidth]{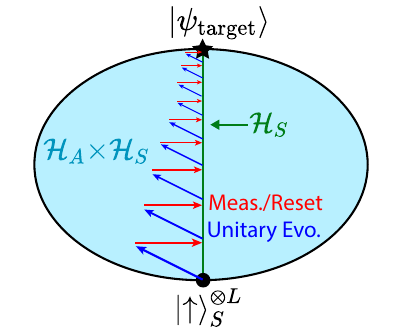}}
    \subfloat[]{\raisebox{0.065\textwidth}{\includegraphics[height=0.17\textwidth]{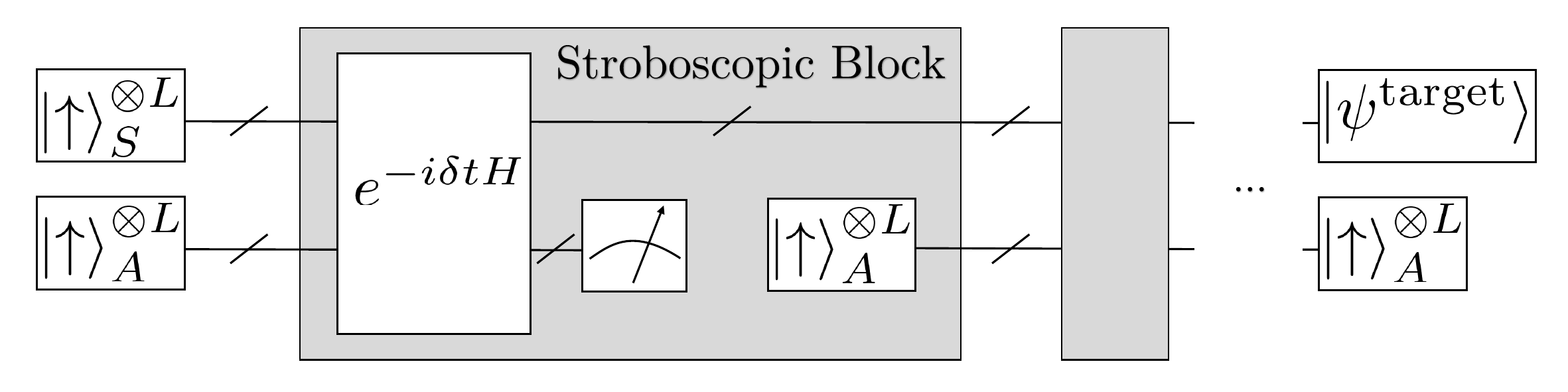}}}
    
    \caption{This figure illustrates the stroboscopic map. Subfigure (a) depicts the quantum systems evolution, where the state is initially $\ket{\uparrow}_S^{\otimes L}$, which resides entirely within the system's Hilbert space (green line). The unitary evolution with the steering Hamiltonian for a duration of $\delta t$ (blue arrows) entangles the system and the ancillas, leading the state to occupy the combined Hilbert space $\mathcal{H}_S \times \mathcal{H}_A$ (light blue surface). Measurement or resetting of the ancillas (red arrows) project back the state into $\mathcal{H}_s$, bringing the system closer to the target state. Upon repeated application of this procedure, the system converges to the target state $\ket{\psi_\text{target}}$. Subfigure (b) represents the same process as a quantum circuit.}
    \label{fig:steering:protocol}
\end{figure*}%

Here, we test this idea on a protocol introduced by Roy \etal  \cite{roy2020measurement} for quantum state steering into states with frustration-free parent Hamiltonians. Specifically, we will be investigating the AKLT state, a spin-1 symmetry-protected topological state (SPT) \cite{verresen2017one, pollmann2012detection} with spin-$\frac{1}{2}$ edge states, is used in this study as a test case for the proposed protocol. This state has potential applications in the study of phases of matter and as a resource for Measurement-Based Quantum Computing \cite{brennen2008measurement}. The protocol is applicable to any state with a frustration-free parent Hamiltonian, like for example the cluster state. In this work, we select the AKLT state for both its simplicity and its potential usefulness.
In a nutshell, we utilize measurements and resets to eliminate undesired portions of the Hilbert space from the available dynamical trajectories of the quantum state.
Thus, we effectively drive the system into the target state of interest.

In order to do so, and as illustrated in Fig.~\ref{fig:steering:protocol}, the system is coupled to ancillas with a time-independent Hamiltonian for a time period of $\delta t$. The protocol exploits the frustration-free nature of the parent Hamiltonian, enabling the writing of local operators that map from locally excited states to locally unexcited states. 
Coupling these local operators to ancilla qubits results in a bias towards the desired non-unitary dynamics if the ancillas are reset periodically.
On a different note to previous works, we determine the convergence time as a function of the resetting time $\delta t$, and investigate the stroboscopic map in regimes beyond the Markovian limit. Note that in the weak-coupling/Markovian limit, our protocol exhibits similarities to the Zeno effect, as rapid successive measurements effectively freeze the state's evolution. However, a key difference arises in stronger coupling regimes, where some evolution is permitted, and the ancillas are reset to their initial states, distinguishing the protocol from the Zeno effect setup.

This study is organized as follows: In Sec.~\ref{sec:stroboscopic_map}, the potential of a stroboscopic map, defined as $\rho(t+\delta t)=\Lambda_{\delta t}\left[\rho(t)\right]$, for preparing desired quantum states is investigated. In Sec.~\ref{sec:AKLT}, it is demonstrated how the protocol can be used to steer into the AKLT state, and a criterion for selecting optimal mapping operators that minimize both entanglement generation and convergence times is proposed, since the choice of mapping operators is non-unique.

In Sec.~\ref{sec:stroboscopic}, full simulations of the system and ancillas are conducted using Matrix Product States (MPS) and quantum trajectories to examine the stroboscopic map in regimes beyond the purely Markovian limit. Furthermore, optimal measurement intervals are determined, and two distinct limits of the system's evolution are identified: the \emph{weak-coupling} and the \emph{strong-coupling} limits. In the weak-coupling limit, the system exhibits Markovian behavior akin to Lindblad dynamics, with an increased convergence speed for larger values of $\delta t$. This behavior persists up to a certain threshold for $\delta t$.

Conversely, in the strong-coupling limit, the convergence rate decreases as the time between successive measurements and resets grows. Remarkably, a broad optimum between these two regimes is uncovered, the position of which can be estimated using a simplified model of commuting mapping operators. In Sec.~\ref{sec:optimal_mapping_operators}, the simplified model of commuting mapping operators is employed to pinpoint optimal mapping operators and to shed further light on the relationship between minimizing the entropy generated by the protocol and the convergence time.

Lastly, the stability of the protocol to dephasing noise and the introduction of a stopping time selection scheme based on information obtained from the ancilla measurements is discussed in Sec.~\ref{sec:errors}.

In conclusion, our study provides valuable insight into the different regimes of state preparation beyond traditional Markovian dynamics and the potential applications of the stroboscopic map for quantum state preparation.


\section{The protocol}

\label{sec:stroboscopic_map}

In this Section, we summarize the general features of our protocol for ancilla-assisted quantum state preparation, as illustrated in Fig.~\ref{fig:steering:protocol}. The idea has its roots in quantum reservoir engineering \cite{Poyatos:1996,Pielawa:2007,Diehl:2008,Verstrate:2009, schirmer2010stabilizing}. The dynamics are designed such as to pull a quantum system into a target state $\rho_0$ as the result of the interplay between coherent and incoherent dynamics. The target state is thus the steady state. Assuming that the system's dynamics can be described by a linear map $\Lambda$, the target state is a fixed point of the map: 
\begin{align}
    \rho_0=\Lambda[\rho_0]\,.
\end{align}
In our work, the system is a many-body system, and the target state is a symmetry-protected topological state. In order to benchmark our analysis we analyze the quantum state preparation of a spin-1 chain in the AKLT state. Below we review the general idea for generating the map $\Lambda$ and then describe the specific implementation for the case here considered. 

\subsection{Time-periodic master equation}

In order to generate $\Lambda$, the system S is coherently coupled to an ancilla $A$. The ancilla $A$ subsequently experiences non-unitary dynamics, characterized by a periodic resetting of its state to a reference state. The composite Hilbert space of the system and ancilla is denoted by $\mathcal H=\mathcal H_S\otimes \mathcal H_A$. The time-dependent master equation governing the density operator $\chi(t)$ in $\mathcal H$ is represented by $\mathcal L(t)$, given as:
\begin{align}
\label{Eq:mEq}
{\mathcal L}(t)\chi(t)=\frac{1}{{\rm i}\hbar}[H,\chi(t)]+\sum_{n}\delta(t-n\delta t){\mathcal K}\chi(t)
\end{align}
In this equation, the Hamiltonian $H$ describes the interaction between the system and the ancilla. The superoperator ${\mathcal K}$, which acts periodically on the ancilla with a period of $\delta t$, is responsible for resetting its state, and assumed to act on a much faster timescale than the dynamics. Master equations such as Eq.~\ref{Eq:mEq} have been explored in the literature to describe the dynamics of masers pumped by beams of atoms \cite{Filipowicz:1986,Slosser:1990,Briegel:1995,Wellens:2000,Pielawa:2010} and, more generally, in order to extend Floquet theory to open quantum systems \cite{Hartmann:2017}. In some cases, the ancilla is reset to different states as a function of $n$ \cite{Wellens:2000}.

In the context of microwave cavity quantum electrodynamics, the operator ${\mathcal K}$ captures the effect of an atom interacting with the maser at short intervals of time. When these interaction times are much shorter than the other time scales of the dynamics, they are approximated by effective kicks occurring at the instants $t_n=n\delta t$ \cite{Briegel:1995,Pielawa:2010}. In our case, instead, the interaction time between the system and the ancilla stretches over a finite interval $\delta t$, in which they become entangled. These dynamics are encompassed by the Hamiltonian
\begin{equation}
    H=H_S+H_A+H_{SA}
\end{equation}
where $H_S$ and $H_A$ denote the Hamiltonian of subsystems S and A, respectively, and $H_{SA}$ their interaction.

The dynamics are generated by Eq.~\ref{Eq:mEq}, with the additional constraint that superoperator $\mathcal K$ leads to an instantaneous resetting of the state of the ancilla to a reference state. The density matrix immediately after the instant of time $t_n$ is given by 
\begin{align}
    \chi(t_n^+)=\rho(t_n)\otimes \varrho^A\,,
\end{align}
where $t_n^{\pm}=\lim_{\epsilon\to 0^+}t_n\pm\epsilon$. In this equation, $\varrho^A$ is the ancilla's reference state, and $\rho(t_n)$ is the system's state obtained by tracing out the ancilla degrees of freedom from the density operator immediately before the resetting:
\begin{align}
    \rho(t_n)={\rm Tr}_A\{\chi(t_n^{-})\}.
\end{align}
The protocol consists of identifying the operator $H$ and the superoperator $\mathcal K$ that leads to the desired map $\Lambda$. 

Some general considerations are in order. We first notice that, for finite periods $\delta t$, the asymptotic state of the system $\rho_{SS}$ is periodic, namely,
$\rho_{SS}(t+\delta t)=\rho_{SS}(t)$. The state $\rho_0$ is the time-average of $\rho_{SS}(t)$ over a period $\delta t$ and the map $\Lambda$ is the one corresponding to the time-averaged dynamics, $\rho_0=\int_0^{\delta t}{\rm d}\tau \rho_{SS}(\tau)/\delta t$. The limit $\delta t\to 0$ leads to a time-independent master equation \cite{Briegel:1995}. This master equation can be reduced to a time-independent Born-Markov master equation for the density matrix of the system $\rho$ when the ancilla-system coupling is sufficiently weak: Assuming that \edit{$J$} is the characteristic frequency characterizing the system-ancilla coupling, then the weak-coupling limit corresponds to \edit{$J\delta t\ll 1$} \cite{Pielawa:2010}. The strong-coupling limit, on the other hand, can be characterized by limit cycles. The latter have been discussed in microwave cavity quantum electrodynamics \cite{Filipowicz:1986,Slosser:1990,Pielawa:2010} and were instrumental in order to prepare Fock states of a microwave resonator \cite{Weidinger:1999,Varcoe:2000}. Limit cycles in the time-periodic master equation have been also proposed for the quantum state preparation of trapped ions \cite{Morigi:1997} and ultracold atoms \cite{Morigi:1998}. In the following, we consider a protocol for preparing a spin-1 chain into the AKLT state and systematically analyze its efficiency as a function of the periodicity $\delta t$, ranging from the weak-coupling to the strong-coupling limit. 

\subsection{Designing the system-ancilla coupling}

Now, consider the density matrix $\chi(t)$. Its stroboscopic evolution is expressed as:
\begin{align}
\chi(t_n^+)={\rm Tr}_A\{{\rm e}^{-iH\delta t}\chi(t_{n-1}^+){\rm e}^{iH\delta t}\}\edit{\otimes}\varrho^A\,.
\end{align}
A significant challenge in quantum reservoir engineering is identifying suitable Hamiltonian operators $H$ to design the target dynamics. For a map $\Lambda$, corresponding to a Lindblad time-independent master equation, one must identify the Kraus operators and construct the associated Hamiltonian. This procedure is similar to unraveling the master equation \cite{Dalibard:1992,Dum:1992}. A convenient approach is outlined in \cite{roy2020measurement}, which applies the von Neumann theory of measurement to develop protocols that steer quantum systems towards pure target states. According to this approach, \edit{$H_S=H_A=0$} and the ancilla-system Hamiltonian adopts the form:

\begin{align}
\label{H:SA}
H_{SA} = J\sum_{l,\alpha}   \edit{M^S_{l,\alpha} \otimes D^{A\dagger}_{l,\alpha}} + \text{h.c}
\end{align}

Here the parameter $J$ is the coupling strength, $D^{A\dagger}_{l,\alpha}$ are local operators that map the ancillas into a state orthogonal to the initial state, and $M^S_{l,\alpha}$ maps the system into the target subspace. More specifically:
\begin{align}
D^{A\dagger}_{l,\alpha} &= \ket{\overline{\Phi^A_{l,\alpha}}} \bra{\Phi^A_{l,\alpha}}\\
M^S_{l,\alpha} &= \ket{\psi^\text{target}_{l,\alpha}} \bra{\psi^\text{undesired}_{l,\alpha}}.
\end{align}
In these equations, $\ket{\Phi^A}$ represents the initial ancilla state, and $\ket{\overline{\Phi^A_{l,\alpha}}}$ are ancilla states chosen to be orthogonal to the initial state $\varrho^A=\ket{\Phi^A} \bra{\Phi^A}$ the ancilla is periodically reset to. The \edit{states} $\ket{\psi^\text{undesired}_{l,\alpha}}$ span the undesired subspace, \edit{while the states} {$\ket{\psi^\text{target}_{l,\alpha}}$ only need to span part of the target subspace}. The index $l$ represents the lattice site where the operator is applied, and $\alpha$ denotes the operator type. The target state is a pure state, $\ket{\psi}_0$, and resides \edit{outside the local subspaces of all the $\ket{\psi^\text{undesired}_{l,\alpha}}$}, rendering it a fixed point of the dynamics ($[H,\rho_0\ket{\Phi^A} \bra{\Phi^A}]=0$) and ensuring convergence. \edit{Note that the $\ket{\psi^\text{target}_{l,\alpha}}$ do not need to span the entire target subspace. For the protocol to converge to the target state, two conditions must be met. First, the mapping operators must steer out of the undesired subspace, so that any state that is in the undesired subspace will be mapped to a state closer to the target state. Second, the mapping operators must be chosen so that their effects at neighboring lattice sites do not disrupt each other, as this can potentially lead to the dynamics being trapped in metastable states.} \edit{We formalize "not disrupting each other" as mapping operators at nearest neighbors having a sufficiently small commutation relation (see~Sec.~\ref{sec:optimal_mapping_operators}). This will be demonstrated in the next sections for the AKLT state, where the $\ket{\psi^\text{target}_{l,\alpha}}$ actually span only a part of the target subspace.}

Let us first consider the weak-coupling limit or Markovian limit, $\edit{J}\delta t\ll 1$. Then, the system's dynamics $\rho(t)=\Tr_A(\chi(t))$ can be described by an effective Lindblad equation \cite{Briegel:1995,Pielawa:2010,roy2020measurement} :
\begin{align}
\partial_t \rho(t) = J \delta t \sum_{l,\alpha}\Bigl( &M^S_{l,\alpha}   \rho(t) M^{S\dagger}_{l,\alpha} \nonumber \\ &- \frac{1}{2}\{ M^{S\dagger}_{l,\alpha} M^S_{l,\alpha}, \rho(t) \}\Bigr).
\label{eq:steering:lindblad_limit}
\end{align}
where time \edit{is now rescaled by $1/J$ to be unitless $Jt \rightarrow t$}.
In this limit, the mapping operators become effective jump operators that map into the target state. This equation suggests, however, that the rate of convergence towards the steady state increases with $\delta t$, and is thus expected to be found in the strong-coupling regime, where Eq.~\ref{eq:steering:lindblad_limit} does not apply.

Generally, determining the optimal period $\delta t$ for the fastest convergence rate is a complex task and often requires numerical approaches \cite{Pielawa:2010}. However, in a specific limit, an explicit derivation of an expression for the fidelity evolution for generic $\delta t$ is possible. This requires constraining the mapping operators to satisfy the commutation relations: $$[M^S_{l,\alpha}, M^{S}_{l',\alpha'}]=0 \quad \wedge \quad [M^S_{l,\alpha}, M^{S\dagger}_{l',\alpha'}]=0 \quad \forall l\neq l'\,.$$ In this case, the fidelity ${\mathcal F}(t)=\bra{\psi_0}\rho(t)\ket{\psi_0}$ evolves according to:
\begin{align}
{\mathcal F}(t+\delta t)=&{\mathcal F}(t)\nonumber\\
&+\sin(\edit{J}\delta t)^2 \sum_{l,\alpha}\bra{\psi_0} M^S_{l,\alpha}\rho(t) M^{S\dagger}_{l,\alpha}\ket{\psi_0}.
\end{align}
This equation demonstrates that selecting $\edit{J}\delta t = \frac{\pi}{2}$ leads to the fastest convergence. Although it is only possible to steer into simple quantum states when using operators that satisfy the commutation relation above, this result offers valuable insights into mapping operators that do not entirely fulfill the commutation relations. Specifically, in Sec.~\ref{sec:optimal_mapping_operators} we derive mapping operators that optimize the protocol efficiency taking this into consideration.

Before concluding this section, it is important to highlight a key observation. While in the weak-coupling limit, the ancilla and system essentially exist in a separable state due to the Markovian nature of the limit \cite{Cubitt:2003}, they become entangled in the strong-coupling limit, revealing the protocol's non-Markovian nature. In this regard, the quantum nature of the ancilla is crucial to the protocol. Furthermore, the reset operation of the ancilla state is effectively a quantum resetting \cite{Roldan:2017,Mukherjee:2018,Yin:2023}.%
\begin{figure}[t]
    \centering
    \includegraphics[width=0.45\textwidth]{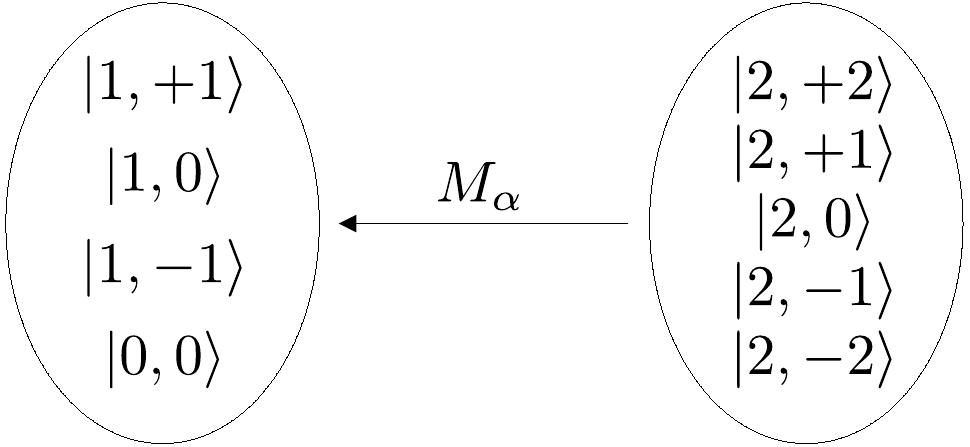}
    \caption{The mapping operators $M_\alpha$ are defined such that all spin states resulting from the combination of two spin-1 sites map from the spin-2 subspace to the spin-(0,1) subspace.}
    \label{fig:steering:mapping_operator_diagramm}
\end{figure}%
\begin{figure*}[t]
    \centering
    \subfloat[]{\label{fig:steering:AKLT_convergence:lindblad}\includegraphics[height=0.3\textwidth]{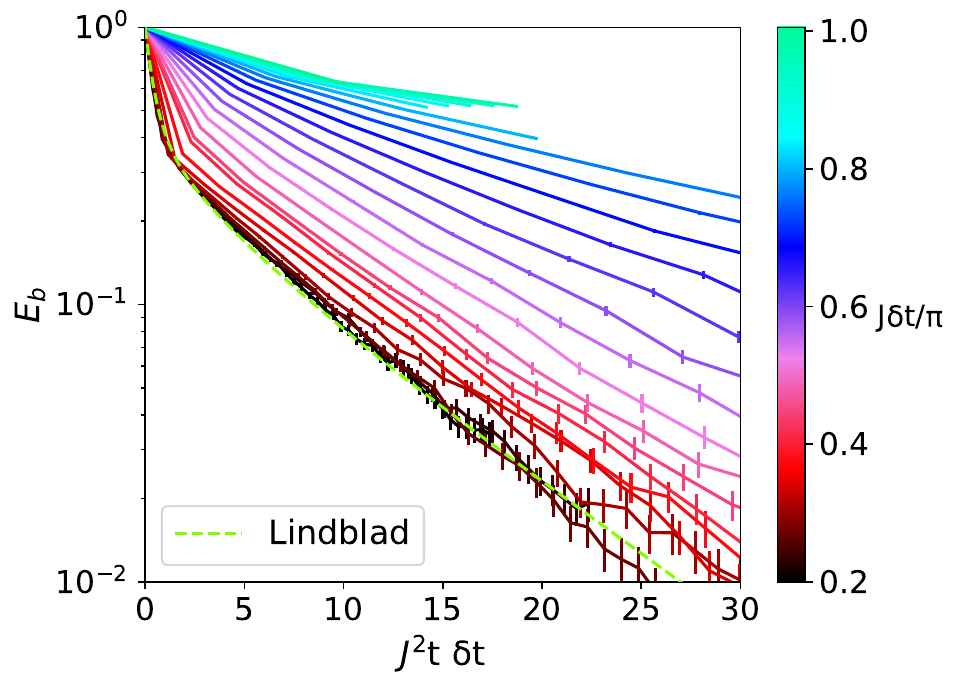}}
    \subfloat[]{\label{fig:steering:AKLT_convergence:periodicity} \includegraphics[height=0.3\textwidth]{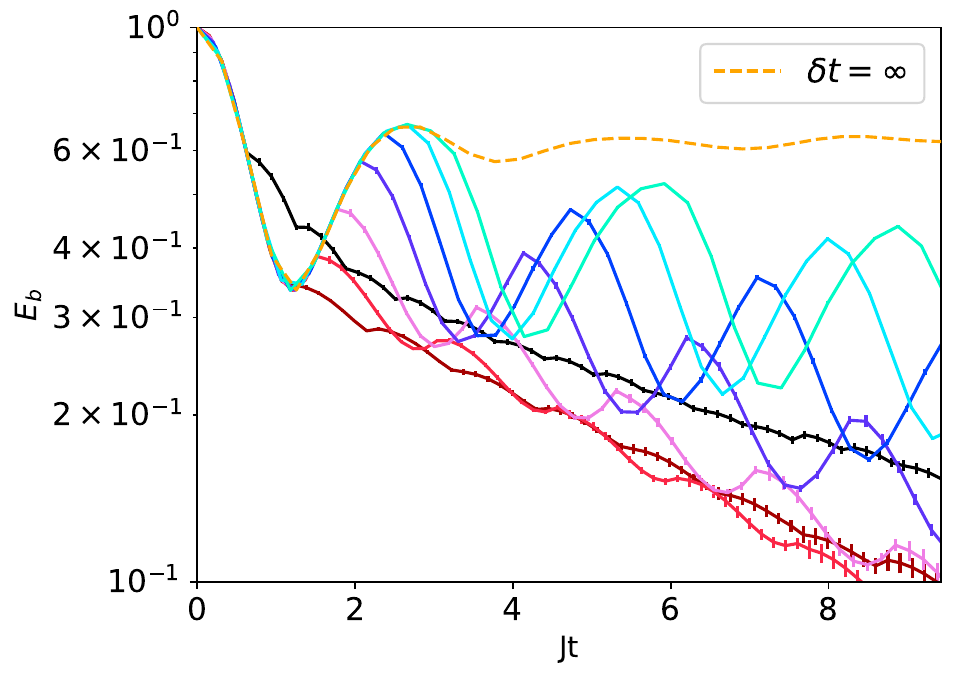}}

    \subfloat[]{\label{fig:steering:AKLT_convergence:energy}\includegraphics[height=0.3\textwidth]{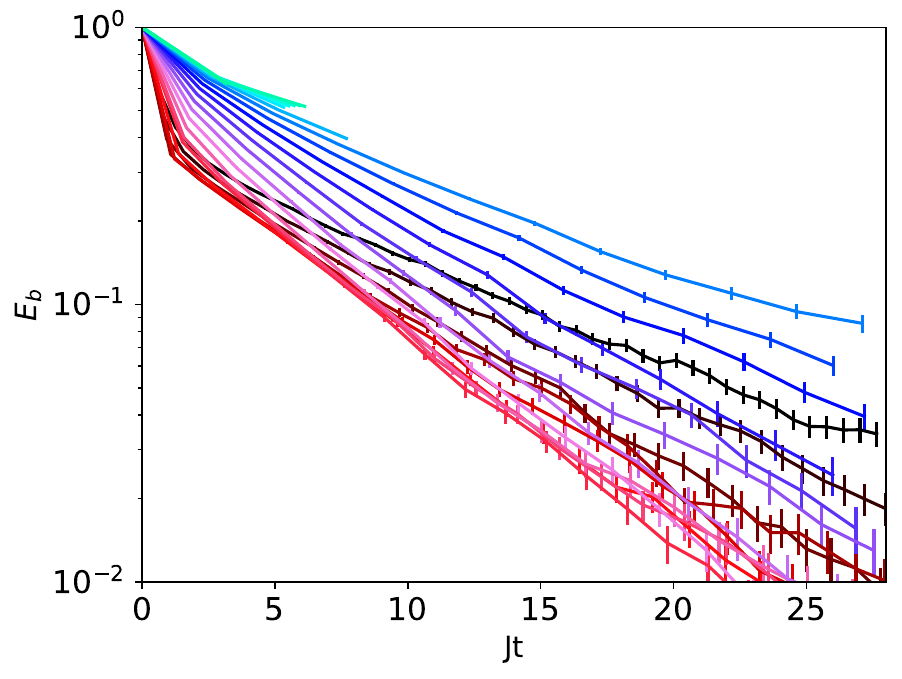}}
    \subfloat[]{\label{fig:steering:AKLT_convergence:fidelity}\includegraphics[height=0.3\textwidth]{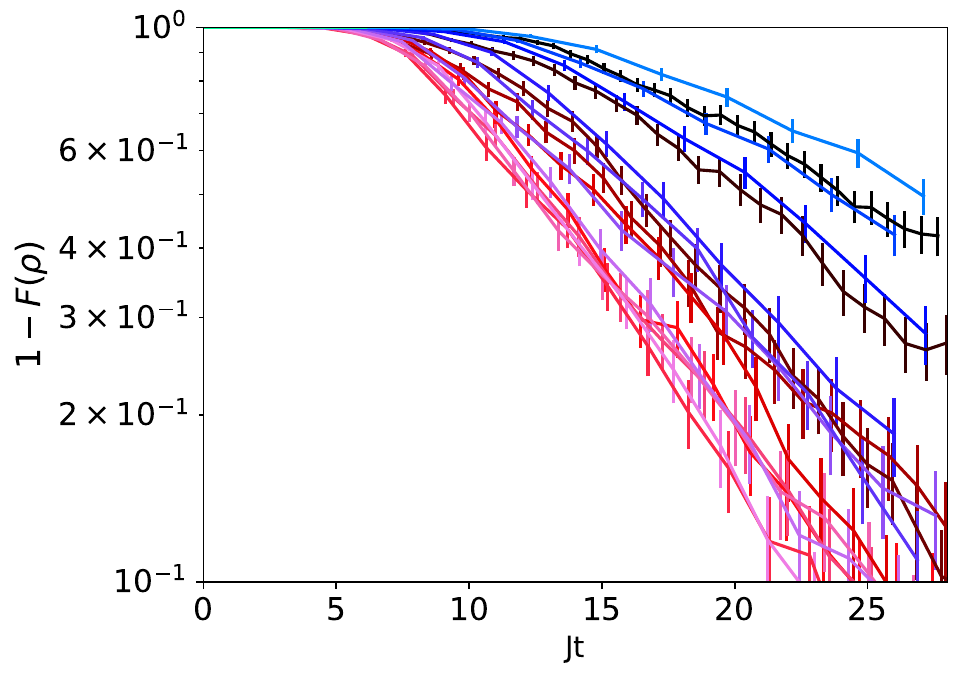}}
    
    \caption{This Figure shows the convergence of the protocol for different measurement intervals $\delta t$ in a system of size $L_s = 15$. Figure a) compares the dynamics obtained from projectively measuring the ancillas with the effective Lindblad dynamics ($\edit{J}\delta t \rightarrow 0$). It is expected that the measurement-induced behavior matches the Lindblad dynamics up to a factor of $\edit{J}\delta t$ in time (Eq.~\ref{eq:steering:lindblad_limit}), and this seems to be approximately the case for $\delta t < \frac{\pi}{4}$. \edit{Note the rescaled x-axis by a factor of $J\delta t$ to make the comparison to the Lindblad dynamics feasible.} Figure b) shows the periodic behavior of the energy per bond $E_b(t)=\frac{1}{L_s-1} \Tr\left[H_\text{AKLT}\rho(t)\right]$.  Figure c) shows the smoothed version of the energy, where no data is displayed when the dynamics cannot be accurately computed. Lastly, Figure d) shows the average infidelity $1 - F(\rho)=1 - \sum_i^4 \bra{\text{AKLT}}_i\rho\ket{\text{AKLT}}_i$ with respect to the four AKLT states.}
    \label{fig:steering:AKLT_convergence}
\end{figure*}%
\section{Mapping Operators For The AKLT State}
\label{sec:AKLT}
The 1D spin-1 Affleck-Kennedy-Lieb-Tasaki (AKLT) state is a relevant example for evaluating the effectiveness of the stroboscopic map in preparing quantum states. The AKLT state is known for its symmetry-protected topological (SPT) \cite{verresen2017one, pollmann2012detection} properties and spin-$\frac{1}{2}$ edge states. It has also been proposed as a resource in measurement-based quantum computing \cite{brennen2008measurement}. The AKLT state can be prepared with the stroboscopic map due to the frustration-free nature of its parent Hamiltonian, which allows for the use of local operators to map excited states to unexcited ones.

The AKLT state is a ground state of the following Hamiltonian:
\begin{align}
    H_\text{AKLT} &= \sum_l P^2(\vec{S}_l + \vec{S}_{l+1})\nonumber\\
    &=\sum_l \left[\frac{1}{2}\vec{S}_l \cdot \vec{S}_{l+1} + \frac{1}{6} \left( \vec{S}_l \cdot \vec{S}_{l+1}  \right)^2 + \frac{1}{3} \right]
\end{align}
Here, \edit{$P^2_{l,l+1}=P^2(\vec{S}_l + \vec{S}_{l+1})$} are operators that project two neighboring spin-1 $\vec{S}_l$ and $\vec{S}_{l+1}$ into the spin-2 subspace. Note that the Hamiltonian was chosen to be dimensionless. Four distinct AKLT states exist in the subspace spanned by the (zero, one) basis, characterized by the property $P^2_{l,l+1}\ket{\text{AKLT}}_i=0 \quad \forall l$.

To prepare the AKLT state using a stroboscopic map, local operators are required to map from the spin-2 subspace to the spin-(0,1) subspace. To accomplish this, a set of operators $M_\alpha$ must be defined such that all states in the spin-2 subspace are mapped to the spin-(0,1) subspace (see~Fig.~\ref{fig:steering:mapping_operator_diagramm}). There are many different choices for the mapping operators $M_\alpha$, but different choices will lead to drastically different convergence speeds. In the main text, we used the mapping operators:
\begin{align}
M_{l,1}^S=\Bigl[&\ket{1,1}\bra{2,2} + \frac{1}{\sqrt{2}}\ket{1,0}\bra{2,0} + \nonumber \\
&\ket{1,-1}\bra{2,-2} \Bigr]_{l, l+1} \nonumber\\
M_{l,2}^S=\Bigl[&\ket{1,1}\bra{2,1} + \frac{1}{\sqrt{2}}\ket{1,0}\bra{2,0} + \nonumber \\ 
&\ket{1,-1}\bra{2,-1}\Bigr]_{l, l+1}.
\label{eq:steering:U}
\end{align}
These operators were chosen based on an ad hoc condition, to try to conserve quantum numbers as much as possible and thus improve experimental realizability. This condition excludes mapping operators that significantly change the quantum numbers, such as mapping from angular momentum 2 to -1, i.e. $\ket{1,-1}\bra{2,2}$. \edit{Note that these mapping operators are not steering into the $\ket{0,0}$ state, this is not an issue as long as they are steering away from all the states in the spin two subspace.}

In the rest of this paper, we will analyze the protocol using the mapping operators in Eq.~\ref{eq:steering:U}. Before concluding this section we note that the choice of these operators is not optimal. The mapping operators, in fact, can be further optimized on the basis of two important considerations.

The first observation to make is that all the states in the spin-2 subspace should be mapped out with equal strength, leading to the imposition of a condition that the mapping operators must sum to the Hamiltonian:
\begin{align}
    H_\text{AKLT} = \sum_l \left( M_{l,1}^{S\dagger}M_{l,1}^{S} + M_{l,2}^{S\dagger}M_{l,2}^{S} \right).
    \label{eq:steering:ham_sum}
\end{align}
In the limit of $\edit{J}\delta t \rightarrow 0$, combining this with Eq.~\ref{eq:steering:lindblad_limit} yields:
\begin{align}
\partial_t \rho(t) = &-\frac{\edit{J}\delta t}{2} \left\{ H_\text{AKLT}, \rho(t) \right\}  \nonumber \\
&+\edit{J}\delta t \sum_{l,\alpha}M^S_{l,\alpha}   \rho(t) M^{S\dagger}_{l,\alpha}\,.
\label{eq:steering:lindblad_limit:ham}
\end{align}
The first term in Eq.~\ref{eq:steering:lindblad_limit:ham} corresponds to imaginary time evolution, while the second term maps the state from the undesired to the desired subspace. When measuring the ancillas, effectively the imaginary time evolution term is applied when the ancillas are measured to be in the initial state $\ket{\Phi^A}$, and the second term is effectively applied when they are found to be in an excited state. Note that the mapping operators in Eq.~\ref{eq:steering:U} satisfy this first condition.

The second consideration involves the commutation relations of the mapping operators at different lattice sites. If the mapping operators satisfy $[M^S_{l,\alpha}, M^{S}_{l',\alpha'}]=0$ and $[M^S_{l,\alpha}, M^{S\dagger}_{l', \alpha'}]=0$ for all $l\neq l'$, the dynamics will follow a sine function, allowing the target state to be reached after only one measurement with $\edit{J}\delta t=\frac{\pi}{2}$. However, it is not possible to choose mapping operators for the AKLT state that fulfill these commutation relations.

Instead, the convergence time can be further optimized by minimizing the commutator between the mapping operators.
The relationship between the commutation relation and convergence time is further explored in Sec.~\ref{sec:optimal_mapping_operators}.

To steer a state into an AKLT state with $L_s$ qutrits, a total of $L=2L_s -1$ qutrits are necessary. The ancillas, which are also spin ones, are initialized/reset to the spin-up state and are evolved according to:
\begin{align}
D^{A\dagger}_{l,1} &= \ket{0}\bra{\uparrow}_{l} \\
D^{A\dagger}_{l,2} &= \ket{\downarrow}\bra{\uparrow}_{l} \\
\ket{\Phi^A} &= \ket{\uparrow\uparrow\uparrow ...}.
\end{align}
\edit{where  $\ket{\downarrow}$,  $\ket{\uparrow}$, and $\ket{0}$ are the eigenstates of the spin-1 $S_z$ operator.}

\section{The Stroboscopic Protocol: results}
\label{sec:stroboscopic}
\begin{figure*}
    \centering
    \subfloat[]{\label{fig:steering:AKLT_performance}\includegraphics[width=0.45\textwidth]{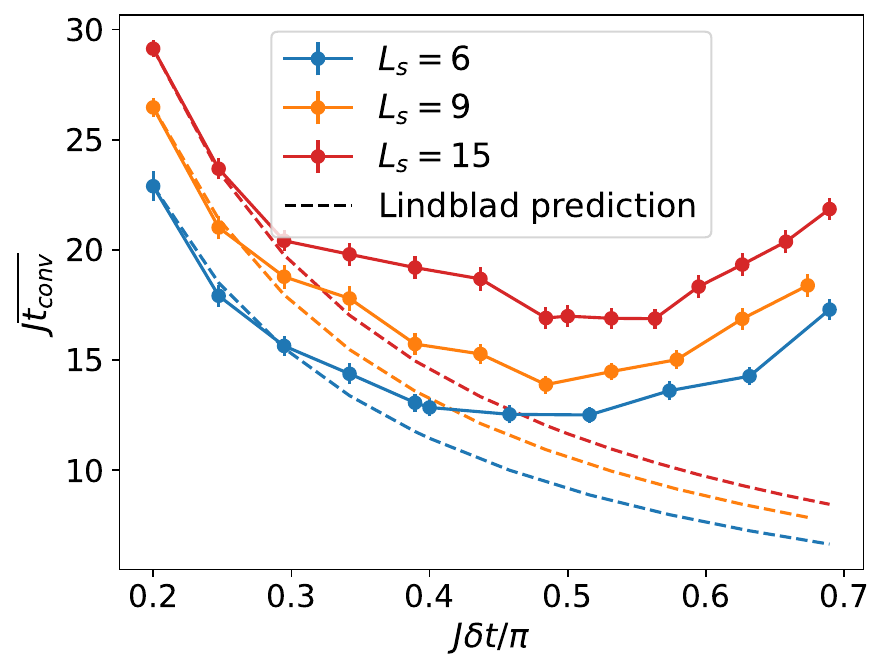}}
    \subfloat[]{\label{fig:steering:AKLT_performance_size}\includegraphics[width=0.45\textwidth]{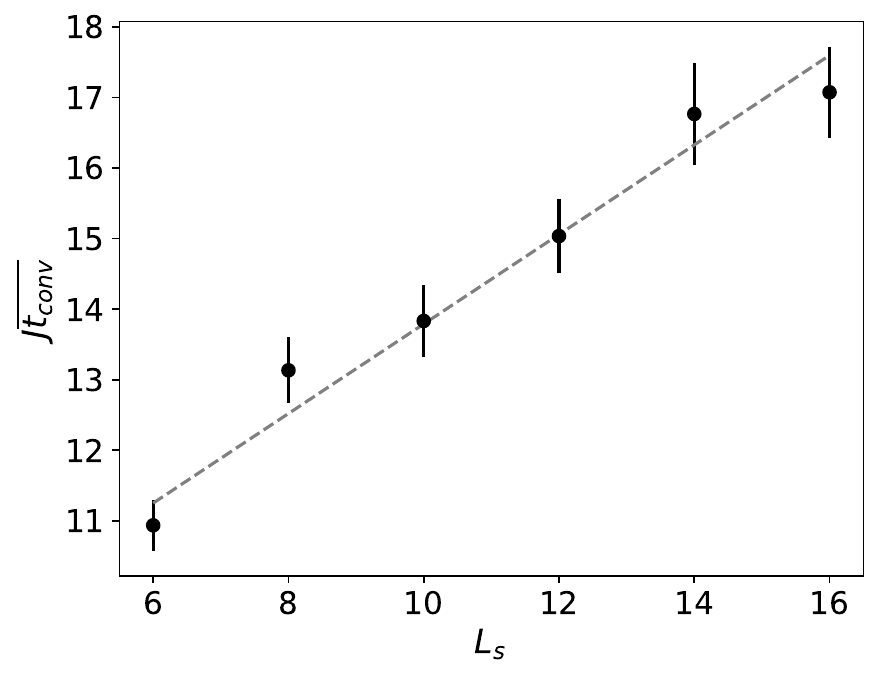}}
    
    \caption{\edit{The figure illustrates the convergence time, denoted as $t_\text{conv}$. This is defined as the time it takes for the ancillas to stop transitioning to states other than their initial spin-up state, effectively stopping their flipping into any excited states.} This marks the beginning of the linear decay of the temperature of the system. $t_\text{conv}$ is plotted as a function of (a) the measurement interval $\delta t$ and (b) the system size $L_s$. In panel (a), we see that $t_\text{conv}$ decreases as $J\delta t$ increases until a minimum is reached around $\frac{\pi}{2}$, while the Lindblad approximation (dashed lines) would predict a decrease of $t_\text{conv}\sim \frac{1}{\delta t}$ for small $\delta t$. Panel (b) shows that $t_\text{conv}$ increases linearly with $L_s$ at a fixed time step of $\edit{J}\delta t = \pi/2$, as indicated by the gray regression line.}
\end{figure*}
In this section, we compare the dynamics of the stroboscopic protocol with those found in the weak-coupling limit governed by Markovian dynamics, which can be described by a Lindblad master equation. Notably, the Markovian limit exhibits characteristics of quantum reservoir engineering, providing a basis for comparing the dynamics of the stroboscopic protocol to those observed in reservoir engineering. Simulation results reveal two distinct limits for the $\delta t$ parameter in the stroboscopic protocol. The first limit, referred to as the weak-coupling limit, adheres to a Lindbladian dynamic, while the second regime, the strong-coupling limit, demonstrates convergence at a slower rate. Furthermore, we identify an intermediate regime that exhibits the fastest convergence and resembles the exceptional case where mapping operators commute. Note that all simulations were performed with the help of the AbelianSymTensor library in Fortran.

The quantum state is initialized as a chain of alternating system and ancilla qutrits in a product state, $\ket{\psi}=\ket{\uparrow \uparrow \dots}$. To time-evolve the state, the Time-Dependent Variational Principle (TDVP) algorithm \cite{haegeman2011time} is employed, with the state represented by a Matrix Product State (MPS). After each time interval $\delta t$, the ancilla qutrits are measured and reset to the spin-up state. This measurement and reset process projects the quantum state closer to the target state. Through repeated iterations and a sufficiently small $\delta t$, the state converges to the target state. This process is simulated using quantum trajectories \cite{daley2014quantum}.

\subsection{Weak-Coupling Limit}
In the weak-coupling limit, where $\edit{J}\delta t < \frac{\pi}{4}$, the dynamics resemble those of Markovian dynamics. \edit{Consequently, it is expected that the convergence speed should be proportional to the measurement interval, and so the convergence time should be inversely proportional} $t_\text{conv} \propto \frac{1}{\delta t}$ (see Eq.~\ref{eq:steering:lindblad_limit}). The Lindblad limit is compared to simulations at finite values of $\delta t$ in Fig.~\ref{fig:steering:AKLT_convergence:lindblad}. For both paradigms, the energy of the AKLT Hamiltonian per bond, $E_b(t)=\frac{1}{L_s-1} \Tr\left[H_\text{AKLT}\rho(t)\right]$, converges to the AKLT state exponentially fast. As anticipated, for small measurement intervals $\edit{J}\delta t < \frac{\pi}{4}$, the evolution at finite $\delta t$ matches the solution of the effective Lindbladian for $\edit{J}\delta t\rightarrow 0$. Outside of the Markovian limit, the energy per bond decays more slowly than predicted by the Lindbladian but still converges to the AKLT state.

To gain a deeper understanding of the energy evolution for finite values of $\delta t$, one can refer to Fig.~\ref{fig:steering:AKLT_convergence:periodicity}. The yellow dashed line illustrates the energy evolution in the absence of measurements, offering insight into the convergence speed for various $\delta t$ values. This line exhibits damped energy oscillations that would continue if measurements were to cease at any point during the protocol. The speed of convergence to the target state is determined by the energy value at which the damped oscillation is interrupted by measurement and reset, initiating a new cycle of damped energy oscillations that will be interrupted once another measurement is performed.

Two limits corresponding to different values of $\delta t$ can be observed in this damped oscillation. In the weak-coupling limit, the energy initially decreases when the Hamiltonian is turned on. In the absence of measurements, the intermediate regime is entered, where the energy follows a sinusoidal pattern until it reaches saturation at a value below the initial energy in the strong-coupling limit. When the ancillas are reset, the sinusoidal energy decrease starts again from the beginning, which is different from the case without measurements. The black line in Fig.~\ref{fig:steering:AKLT_convergence:periodicity}, representing the weak-coupling limit with a resetting interval of $\edit{J}\delta t = 0.2\pi$, behaves similarly to the Lindbladian as discussed earlier. Since the measurement interval for the black line is fast, it is not affected by the sinusoidal nature of the measurement-free evolution. In contrast, larger values of $\delta t$ in the intermediate regime $0.4\pi<\edit{J}\delta t <0.6\pi$ can exploit the sinusoidal structure to minimize the energy in each step of the process. However, in the strong-coupling limit $0.6\pi<\edit{J}\delta t$, larger values of $\delta t$, measured after the sinusoidal structure has already started to increase, resulting in slower convergence times.

Fig.~\ref{fig:steering:AKLT_convergence:energy},\subref*{fig:steering:AKLT_convergence:fidelity} further demonstrates the convergence behavior of the energy/infidelity for different values of $\delta t$. Both quantities show the fastest convergence when in the intermediate regime with $\edit{J}\delta t_\text{opt} \approx \frac{\pi}{2}$. This is consistent with the observations made in the previous paragraph.

It should be noted that for values of $\delta t$ in the strong-coupling limit, accurate simulations of the system are not possible due to technical limitations. As measurements become less frequent, the quantum state accumulates more entanglement, which cannot be adequately represented by an MPS (see~Sec.~\ref{sec:large_deltat_t}). As a result, the lines in Fig.~\ref{fig:steering:AKLT_convergence:energy},\subref*{fig:steering:AKLT_convergence:fidelity} are discontinued once these technical limitations arise.

\subsection{Convergence times}
\label{sec:convergence_time}
To further examine the convergence behavior, we define the convergence time as the moment when the ancillas cease to flip into an \edit{excited state (i.e., any state other than the initial state $\ket{\uparrow}$) and instead stay in their initial state}. This event occurs when the system's excitations are insufficient to induce transitions excitations in the ancillas, allowing only the imaginary time evolution term in Eq.~\ref{eq:steering:lindblad_limit:ham} to act on the state. As a result, the cessation of ancilla flipping corresponds to the cooling of the state, leading to a reduction in the inverse temperature of the state, denoted by $\Delta \beta(t) = t \cdot \edit{J}\delta t$. Once the state is considered converged, an exponential decrease in energy, characterized by $E(t)\propto e^{-\edit{J} t\cdot \delta t \Delta E}$, where $\Delta E$ is the energy gap of the AKLT Hamiltonian, is observed (see~App.~\ref{appendix:imaginary_time_evolution}).

Fig.~\ref{fig:steering:AKLT_performance} displays the convergence time for the system as a function of $\delta t$ for different system sizes. The optimal value of $\delta t$ is found to be around $\edit{J}\delta t_\text{opt}\approx\frac{\pi}{2}$ for all sizes, corresponding to the intermediate regime where the sinusoidal pattern of the energy evolution enables faster convergence. In contrast, smaller values of $\delta t$ in the weak-coupling limit and larger values in the strong-coupling limit result in slower convergence times due to not exploiting the sinusoidal structure. It is evident that in the weak-coupling limit, the procedure will always converge with a convergence time proportional to $\frac{1}{\delta t}$ (see the dotted lines in Fig.~\ref{fig:steering:AKLT_performance}), as the Lindbladian guarantees it. However, the possibility of convergence in the strong-coupling limit remains uncertain and will be further explored in the following section.

The energy evolution behavior in the intermediate regime can be further understood by considering the case where the mapping operators commute, $[M_{l,\alpha}^{S},M_{l+1,\alpha'}^{S}]=0$ (see~Sec.~\ref{sec:stroboscopic_map}). In this case, the optimal value for $\delta t$ is $\edit{J}\delta t_\text{opt} = \frac{\pi}{2}$, and the energy fluctuates periodically with a sinusoidal pattern. This observation aligns with those made in the intermediate regime. However, if the mapping operators do not commute, the sinusoidal behavior may be altered, leading to slightly different convergence times in the intermediate regime and a complete departure from the commuting case when entering the strong-coupling limit. This commutation relation plays a significant role in finding the optimal mapping operators (see~Sec.~\ref{sec:optimal_mapping_operators}).
 
The system's convergence time, analyzed in Fig.~\ref{fig:steering:AKLT_performance_size} as a function of the system size, exhibits a linear increase with the size of the chain. This observation is attributed to the delocalized nature of the AKLT state, featuring two spin-half edge states. Preparing the AKLT state requires entangling the two edges of the chain, which, for a local protocol, scales linearly with the system size at best. Although protocols employing non-local feedback from measurement results, as demonstrated in \cite{smith2022deterministic}, can achieve non-scaling convergence speeds, these methods have limitations, such as the inclusion of non-local measurement-based feedback, which is not yet widely accessible on current quantum platforms. In light of this, we focus on a minimalist solution for stabilizing these topological states in the NISQ era.

\subsection{Strong-Coupling limit}
\label{sec:large_deltat_t}
\begin{figure*}[t]
    \centering
    \subfloat[]{\label{fig:steering:AKLT_entanglement:evo}\includegraphics[height=0.28\textwidth]{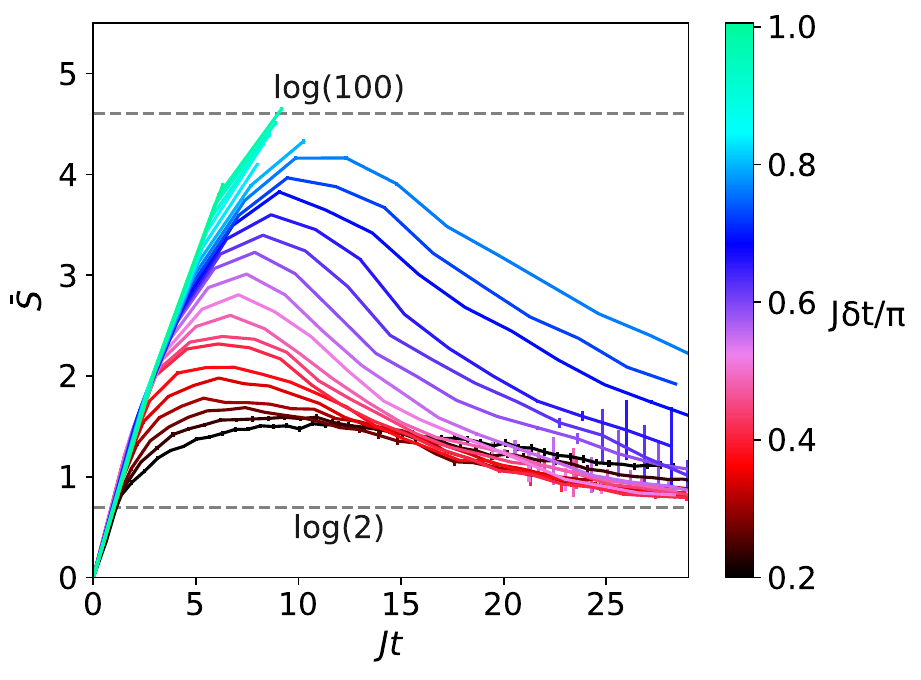}}
    \subfloat[]{\label{fig:steering:AKLT_entanglement:Smax}\includegraphics[height=0.28\textwidth]{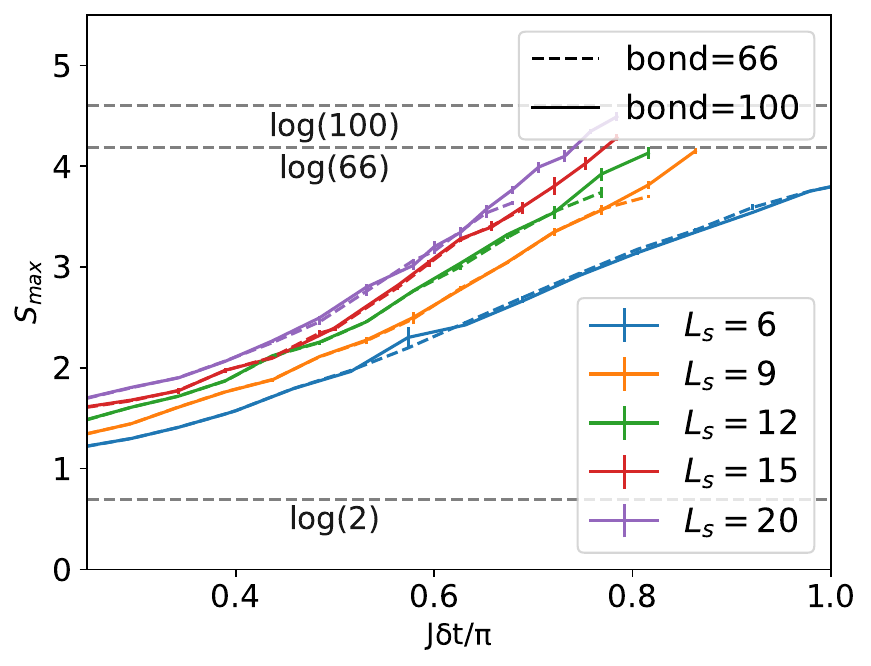}}
    
    \subfloat[]{\label{fig:steering:AKLT_entanglement:tmax}\includegraphics[height=0.25\textwidth]{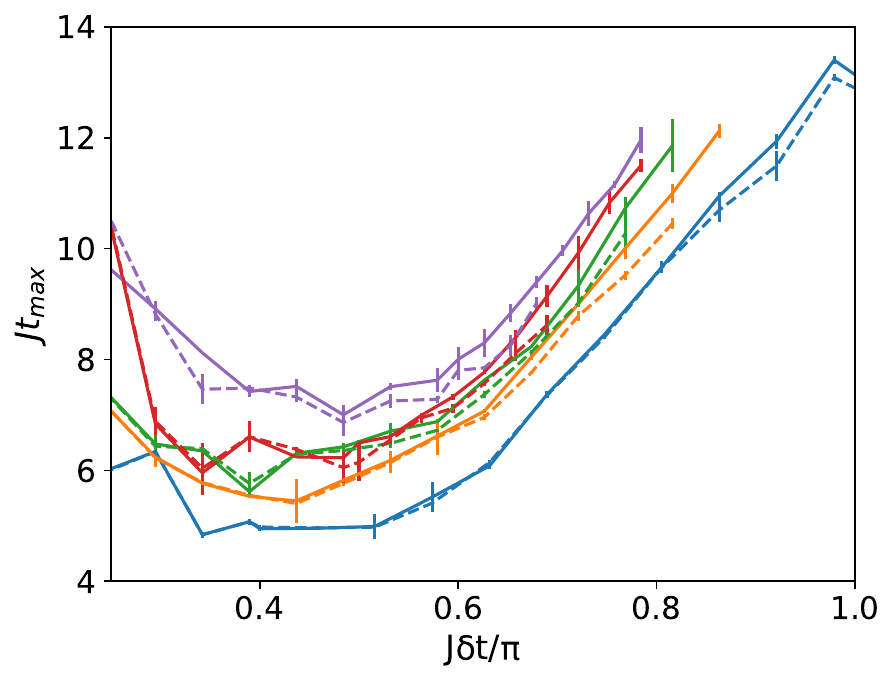}}
    \subfloat[]{\label{fig:steering:AKLT_entanglement:tconv_tmax}\includegraphics[height=0.25\textwidth]{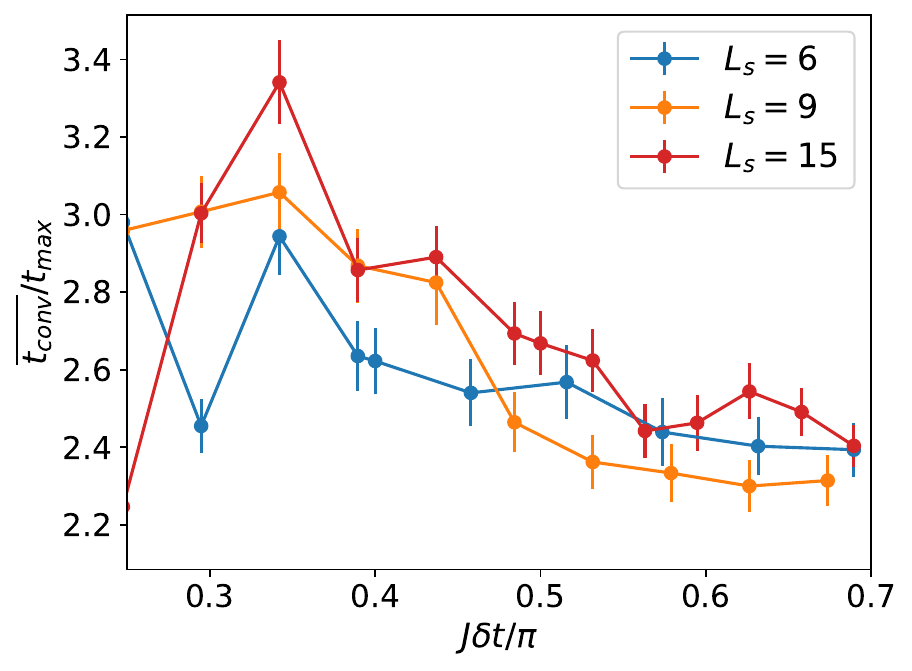}}
    \subfloat[]{\label{fig:steering:AKLT_entanglement:smax-tmax}\includegraphics[height=0.25\textwidth]{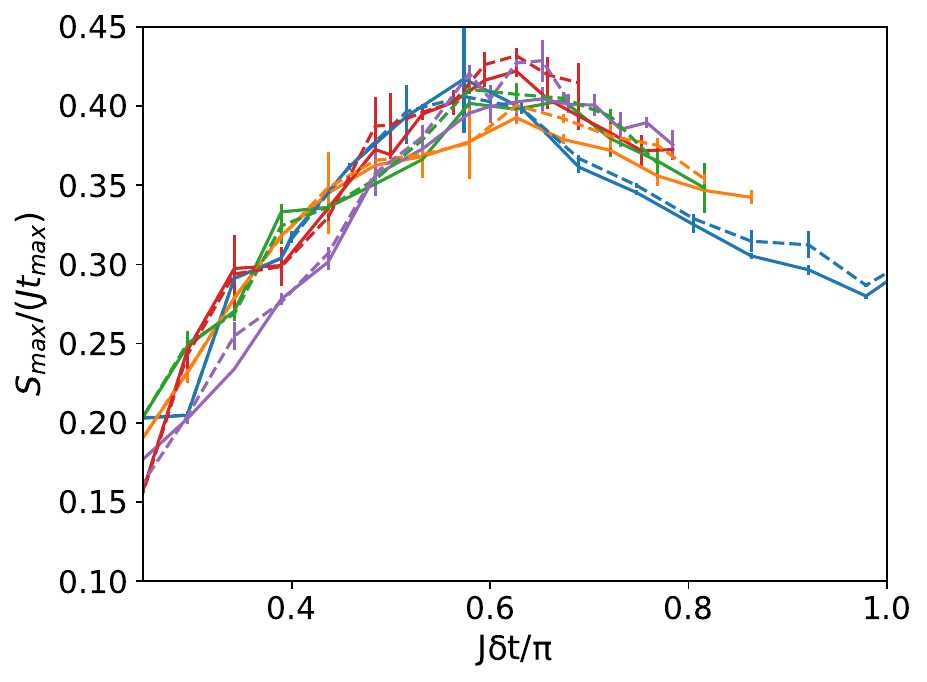}}
    
    \caption{(a) Shows the average extrapolated entanglement $\bar{S}$ for different measurement intervals $\delta t$ at a system size of $L_s=15$. Simulations at large $\delta t$ produce more entanglement than can be accurately represented with a bond dimension of 100 and are thus discontinued when they no longer match simulations with a lower bond dimension. (b) Presents the maximum entanglement reached for different $\delta t$, which are effectively the maxima of Fig. (a)  for simulations with bond dimensions of 66 (dotted line) and 100 (continuous line) (c) Shows the time required to reach the maximal entanglement. (d) Compares the convergence time to the time taken to reach the maximal entropy. (e) Displays the entanglement buildup rate to the maximum $S_\text{max}/t_\text{max}$. }
    \label{fig:steering:AKLT_entanglement}
\end{figure*}%

In the previous section, the strong-coupling limit was characterized by values of $\delta t$ at which the system's energy ceases to display a sinusoidal pattern and instead approaches a constant energy value during each iteration of the protocol. This section focuses on examining the system's behavior under these values of $\delta t$, with particular emphasis on the role of entanglement entropy. Entanglement entropy is computed by partitioning a quantum system into two parts, A and B, and determining the von Neumann entropy of the reduced density matrix for part A. This measure quantifies the extent of quantum correlations between subsystems and is essential for investigating properties of quantum many-body systems, such as phase transitions and entanglement scaling. In this study, attention is directed toward the average entanglement entropy across the pure states in different trajectories characterized by the probabilistic ancilla measurement outcomes.

It is a priori unclear whether the protocol will continue to converge to the AKLT state for larger time intervals, $\delta t$, between measurements. Fig.~\ref{fig:steering:AKLT_convergence:energy} demonstrates that convergence decelerates as $\delta t$ increases, but it remains uncertain whether the protocol will cease converging at a finite $\delta t$. In the absence of measurements, i.e., $\delta t \rightarrow \infty$, the system is expected not to converge to the AKLT state. Moreover, in the no-measurement scenario, the entanglement of the quantum system is anticipated to grow ballistically until it reaches the system's maximal entropy capacity, $\frac{L}{2}\log(3)$. This section investigates if there exists a finite value of the measurement time interval $\delta t$, at which measuring and resetting the ancillas fail to reduce the system's entropy, resulting in exponentially scaling convergence times with system size, as observed in the study of the Measurement Induced Entanglement Phase Transition (MIEPT).

Models featuring MIEPT typically have two different elements: a measurement element that reduces entropy and a Hamiltonian evolution or random circuit element that increases entropy. In our protocol, the entropy-reducing element is the reset of the ancillas, and the entropy-increasing element is the Hamiltonian evolution. In systems exhibiting MIEPT, measurements can counterbalance entanglement growth induced by Hamiltonian evolution up to a certain ratio between the two elements. In such systems, if an insufficient number of measurements are performed, an entanglement phase transition may arise, leading to entanglement growth.

Should a phase transition manifest in the stroboscopic protocol, it would likely materialize as a deviation in the entanglement growth, potentially impeding convergence to the target state. Nonetheless, our analysis employing the entanglement-extrapolation technique outlined in App.~\ref{appendix:entanglement_extrapolation} suggests that this model does not exhibit a phase transition. Instead, it displays characteristics of a volume law phase even for small values of $\delta t$.

The system's entanglement initially increases from its initial value, reaching a maximum at a time $t_\text{max}$ and an entanglement entropy $S_\text{max}$ (see Fig.~\ref{fig:steering:AKLT_entanglement:evo}). Subsequently, the entropy decreases until it reaches the entanglement entropy of the AKLT state. The AKLT state's entropy lies between $\log(2)$ and $\log(4)$, depending on the entanglement degree of the two edge states. This pattern is observed for all values of $\delta t$, with the sole difference being that larger values of $\delta t$ yield higher values of $S_\text{max}$. Our aim is to ascertain whether there exists a value of $\delta t$ for which the system's entanglement begins to increase ballistically. This effect would cause $S_\text{max}$ to increase to $\frac{L}{2}\log(3)$ and would exponentially slow down convergence to the target state. The increase in entropy would also hinder our ability to accurately simulate the system, as the maximum entanglement entropy that an MPS can represent is constrained by the bond dimension to $S \leq \log(\text{bond})$. Nevertheless, this limitation is partially alleviated by our entanglement-extrapolation technique, detailed in App.~\ref{appendix:entanglement_extrapolation}.

Fig~\ref{fig:steering:AKLT_entanglement:Smax} displays the maximal entanglement entropy reached during the evolution, $S_\text{max}$, for various system sizes, $L_s$. The results reveal that $S_\text{max}$ approaches the entropy of the AKLT state as the time step $\delta t$ nears zero and increases linearly for large $\delta t$. As the system size expands, $S_\text{max}$ grows approximately linearly, indicating a volume law phase. Additionally, the time required to reach $S_\text{max}$, as depicted in Fig.~\ref{fig:steering:AKLT_entanglement:tmax}, appears to increase linearly with large $\delta t$. Importantly, no lattice size-dependent discontinuity is observed in either $t_\text{max}$, $S_\text{max}$, or their derivatives, suggesting the absence of a phase transition. Intriguingly, even without biasing the dynamics by resetting the ancillas, the entanglement entropy attains a stationary value similar to $S_\text{max}$ observed in the reset-and-measure protocol. The only distinction to the data shown here is that the entropy does not decrease to the target state because the dynamics are no longer biased without the reset. Nonetheless, even in this modified protocol, there is no evidence of a phase transition.

Remarkably, the time required to reach maximal entanglement exhibits a behavior similar to the convergence time, with a minimum around $\edit{J}\delta t = \frac{\pi}{2}$. For large $\delta t$, $t_\text{max}$ can be accurately calculated, as shorter simulations are necessary. The same linear behavior is observed for $t_\text{max}$ as for the convergence time $t_\text{conv}$ when $\delta t$ is large. Specifically, $t_\text{conv}$ appears to be two to three times larger than $t_\text{max}$ (see~Fig.~\ref{fig:steering:AKLT_entanglement:tconv_tmax}). Since $t_\text{max}$ does not diverge at any point, it is plausible that $t_\text{conv}$ will not diverge either. If this trend persists for larger $\delta t$, it would imply that the protocol converges after a fixed number of measurements for larger $\delta t$.

The analysis of the quantity $\frac{S_\text{max}}{J t_\text{max}}$, illustrated in Fig.~\ref{fig:steering:AKLT_entanglement:smax-tmax}, provides valuable insights. This quantity serves multiple purposes: it reflects the average rate at which the system attains maximal entanglement entropy, and it remains invariant with respect to lattice size, suggesting the absence of a phase transition. Initially, this quantity increases due to the minimum at $\edit{J}\delta t=\frac{\pi}{2}$ in $t_\text{max}$, but it later converges to the ratio of the corresponding slopes of the linear functions $S_\text{max}$ and $t_\text{max}$ in the strong-coupling limit. Examining this quantity further corroborates the conclusion that no phase transition exists in the system, thus emphasizing the protocol's reliability in converging to the target state.

Simulations were performed using bond dimensions of 66 and 100 to validate the findings. Owing to the necessity of executing 128 distinct trajectories for each parameter set, only small bond dimensions were employed. In particular, 13,000 simulations were required to obtain the data presented in Fig.~\ref{fig:steering:AKLT_entanglement:Smax}.

\section{Optimal Mapping Operators}
\label{sec:optimal_mapping_operators}
\begin{figure*}
    \center
    \subfloat[]{\label{fig:steering:optimal_map:energy}\includegraphics[height=0.23\textwidth]{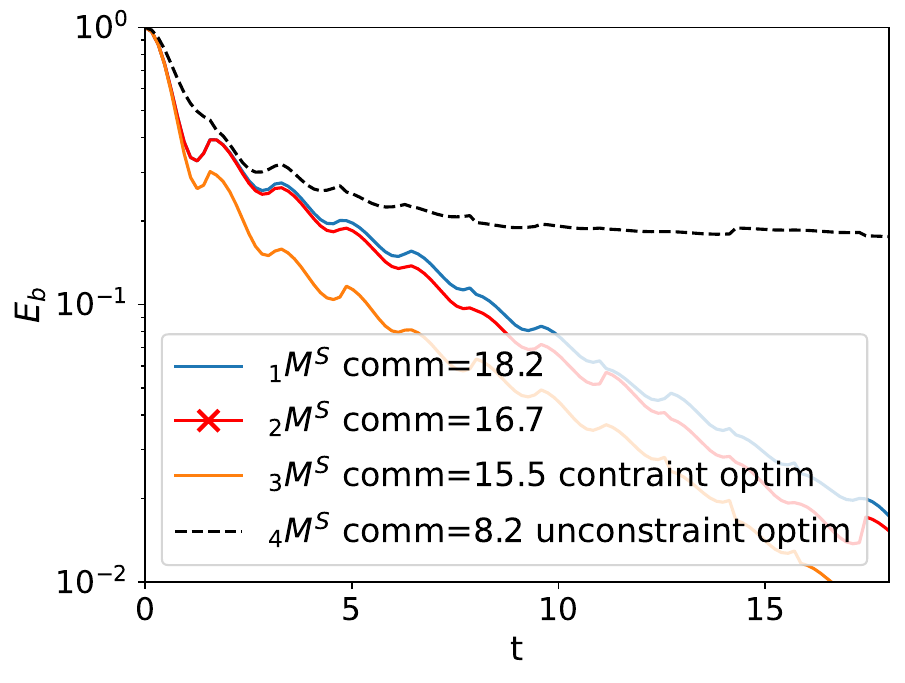}}\subfloat[]{\label{fig:steering:optimal_map:entropy}\includegraphics[height=0.23\textwidth]{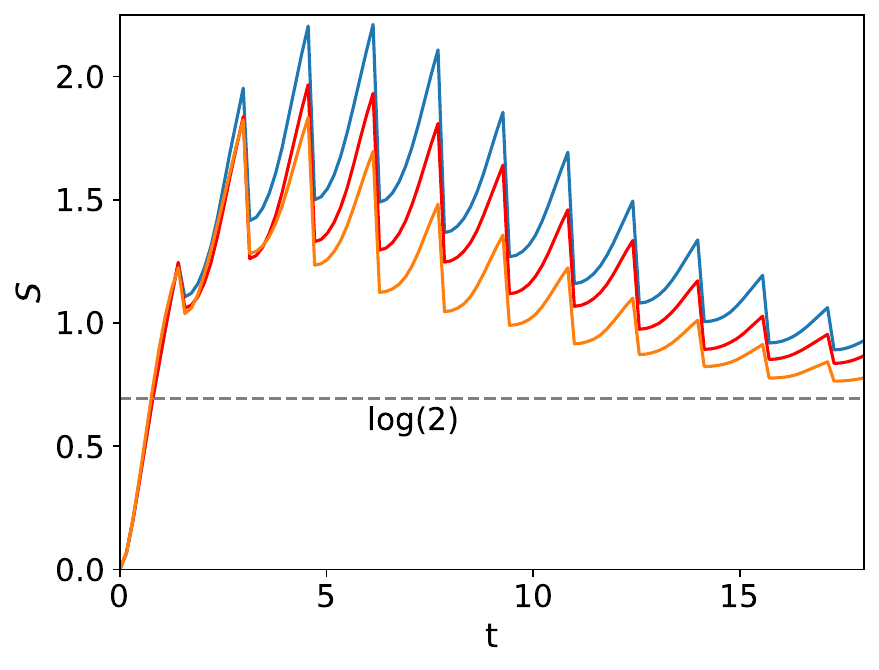}}
    \subfloat[]{\label{fig:steering:optimal_map:tconv}\includegraphics[height=0.23\textwidth]{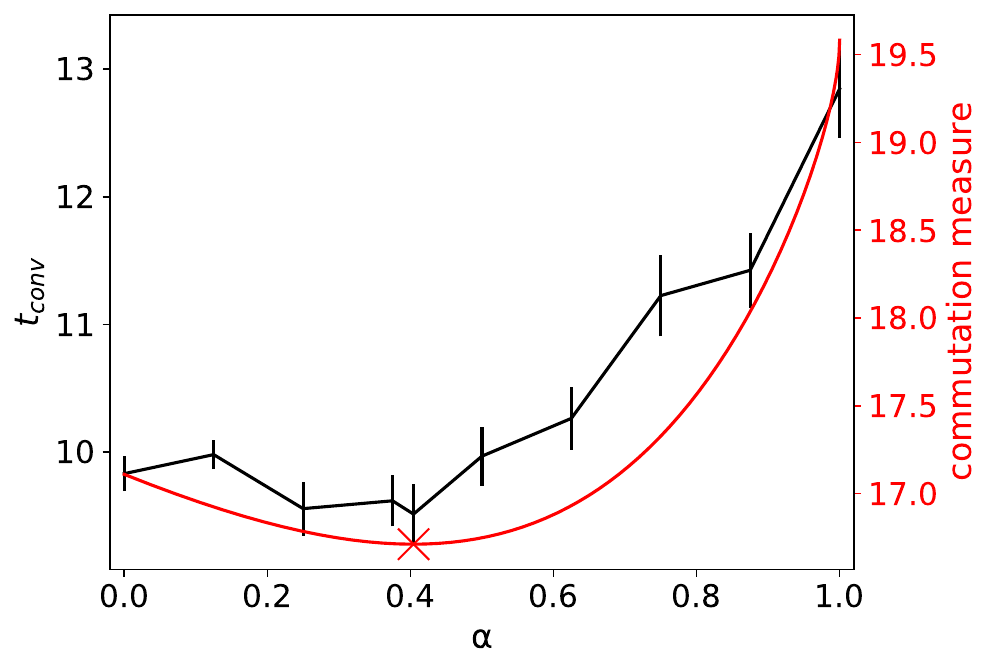}}
    
    \caption{This figure shows the convergence behavior of different mapping operators at a system size of $L=10$. (a) shows the energy per bond for four different mapping operators: the (1) blue curve represents the mapping operators used in the rest of the paper, the (2) \edit{red} curve represents an improved version of the mapping operators, the (3) \edit{yellow} curve represents mapping operators optimized to minimize the commutation measure, and the (4) \edit{gray dotted} curve represents mapping operators optimized to minimize the commutation measure while not enforcing the equation (Eq.~\ref{eq:steering:ham_sum}). (b) shows the entanglement entropy for the different mapping operators. (c) shows the convergence time and the commutation measure for mapping operators $_2M(\alpha)$ for different values of $\alpha$. The minimum value for both measures is observed at $\alpha=0.404$, marked by a red cross in the legend of sub-figure (a) and in sub-figure (c).}
    \label{fig:steering:optimal_map}
\end{figure*}

In this section, we analyze the influence of the choice of mapping operators $M_\alpha$ on the dynamics and convergence time required to prepare a target state. To ensure that the mapping operators map into the AKLT state, they must satisfy two conditions: first, they must map away from the five states in the spin-2 subspace and into the spin-(0,1) subspace (see~Fig.~\ref{fig:steering:mapping_operator_diagramm}); second, they must contribute to the Hamiltonian as outlined in Eq.~\ref{eq:steering:ham_sum}.

If the mapping operators satisfy the commutation relations $[M^S_{l,\alpha}, M^{S}_{l',\alpha'}]=0$ and $[M^S_{l,\alpha}, M^{S\dagger}_{l', \alpha'}]=0$ for all $l\neq l'$, the dynamics exhibit a sinusoidal pattern, and the target state can be reached after a single measurement at $\edit{J}\delta t=\frac{\pi}{2}$ time, as discussed in Sec.~\ref{sec:stroboscopic_map}. In practice, however, it is not feasible to satisfy these conditions for complex target states such as the AKLT state.

If the commutation relations are not satisfied, the operators interfere with each other, leading to the formation of undesired entanglement in parts of the system orthogonal to the target state. The relationship between entanglement growth and convergence times is explored in the previous section.

In order to minimize entanglement generation and thus reduce convergence times, the goal is to approximate the commutation relation as closely as possible. This goal is achieved by minimizing the commutation measure:
\begin{align}
\text{comm}(M)=\sum_{l\neq l',\alpha,\alpha'}&\Vert[M_{l,\alpha}, M_{l',\alpha'}]\Vert_2\nonumber\\
&+\Vert[M_{l,\alpha}, M_{l',\alpha'}^\dag]\Vert_2.
\end{align}
For two nearest-neighbor mapping operators that are identical for all lattice sites, the measure simplifies to:
\begin{align}
\text{comm}(M)=&\Vert[M_{l,1}, M_{l+1,1}]\Vert_2 +
\Vert[M_{l,1}, M_{l+1,2}]\Vert_2 \nonumber \\
+ &\Vert[M_{l,2}, M_{l+1,2}]\Vert_2 + 
\Vert[M_{l,1}, M_{l+1,1}^\dag]\Vert_2 \nonumber \\
+&\Vert[M_{l,1}, M_{l+1,2}^\dag]\Vert_2 +
\Vert[M_{l,2}, M_{l+1,2}^\dag]\Vert_2 .
\end{align}
Note that all mapping operators are also constrained by Eq.~\ref{eq:steering:ham_sum}.

In the previous sections, the mapping operators:
\begin{align}
_1M_{l,1}^S&=\left[\ket{1,1}\bra{2,2} + \frac{1}{\sqrt{2} }\ket{1,0}\bra{2,0}+ \ket{1,-1}\bra{2,-2} \right]_{l, l+1} \nonumber\\
_1M_{l,2}^S&=\left[\ket{1,1}\bra{2,1} + \frac{1}{\sqrt{2} }\ket{1,0}\bra{2,0} + \ket{1,-1}\bra{2,-1}\right]_{l, l+1}
\label{eq:steering:U1}
\end{align}
have been used. These mapping operators have a commutation measure of $\text{comm}(_1M)=18.2$ which is higher than the other mapping operators we will introduce in this section. The $_1M$ operators were chosen for symmetry reasons before the relationship between the commutation measure and the convergence time was found.

The effectiveness of the commutation measure as a criterion for selecting optimal mapping operators is demonstrated by analyzing a parameterized set of operators:
\begin{align}
_2M_{l,1}^S(\alpha)=\bigg[&\ket{1,1}\bra{2,2} + \alpha\ket{0,0}\bra{2,-1} \nonumber\\
&+ \ket{1,-1}\bra{2,-2} \bigg]_{l, l+1} \nonumber\\
_2M_{l,2}^S(\alpha)=\bigg[&\ket{1,1}\bra{2,1} + \sqrt{1-\alpha^2}\ket{1,-1}\bra{2,-1}\nonumber\\
&+\ket{1,0}\bra{2,0}\bigg]_{l, l+1} .
\label{eq:steering:U2}
\end{align}
Fig.~\ref{fig:steering:optimal_map:tconv} illustrates the relationship between the convergence time and commutation measure for different values of $\alpha$. Both measures exhibit a U-shaped pattern, with a minimum at $\alpha=0.404$. This observation suggests that the commutation measure can serve as a reliable indicator for selecting optimal mapping operators.

To optimize the mapping operators, we employed gradient descent techniques to minimize the commutation measure. Each mapping operator was parameterized using a $4\times5$ matrix, where each matrix element maps a state from the spin-2 subspace to the spin-(0,1) subspace.

The resulting optimized operators were minimized to a commutation measure of $\text{comm}(_3M)=15.5$ with the form:
\begin{align}
\alpha=0&.8482 \nonumber\\
_3M_{l,1}^S=\bigg[&\ket{1,1}\bra{2,2} + \ket{1,0}\bra{2,1}\alpha \nonumber\\
&+ \ket{0,0}\bra{2,1}\sqrt{1-\alpha^2} \nonumber\\&+ \ket{1,-1}\bra{2,0}\frac{1}{\sqrt{2}}\bigg]_{l, l+1} \nonumber\\
_3M_{l,2}^S=\bigg[&\ket{1,-1}\bra{2,-2} + \ket{1,0}\bra{2,-1}\alpha \nonumber\\
&+ \ket{0,0}\bra{2,-1}\sqrt{1-\alpha^2}\nonumber\\&+ \ket{1,1}\bra{2,0}\frac{1}{\sqrt{2}}\bigg]_{l, l+1}.
\label{eq:steering:U3}
\end{align}

The table below presents the results for convergence time and commutation measures associated with the three aforementioned mapping operators.

\begin{table}[H]
\centering
\begin{tabular}{|l|l|l|}
 &  comm & $t_\text{conv}$ \\\hline
 $_1M$& 18.2 & 10.94 \\
 $_2M$& 16.7 & 9.51 \\
 $_3M$& 15.5 & 7.66
\end{tabular}
\end{table}

A significant observation is the correlation between the decrease in commutation measure and the decrease in convergence time. The energy evolution for these mapping operators can be examined in Fig.~\ref{fig:steering:optimal_map:energy}. Furthermore, Fig.~\ref{fig:steering:optimal_map:entropy} shows that mapping operators with lower commutation measures produce less undesired entanglement, as hypothesized.

It is worth mentioning that the causal relationship between the commutation relation and convergence time was established only after most of the computationally expensive simulations were completed. In the previous section, the mapping operators in Eq.~\ref{eq:steering:U1} were used. They perform approximately 20\% worse than the fully optimized ones Eq.~\ref{eq:steering:U1} but, no qualitative change between the time evolution was found. Consequently, the original choice of mapping operators was retained to conserve computational resources.

An important observation is that not enforcing the Hamiltonian constraint, as shown in Eq.~\ref{eq:steering:ham_sum}, allows for smaller commutation measures. However, the resulting dynamics exhibit slow convergence, as shown by the black dotted line in Fig.~\ref{fig:steering:optimal_map:energy}. This slow convergence is due to the weak mapping of certain states in the spin-2 subspace to the spin-(0,1) subspace. Such cases occur when the mapping operators sum to an operator with eigenvalues unequal to one for eigenvectors in the spin-2 subspace. It should be noted that the choice of optimal mapping operators is not unique. Several alternatives with comparably small commutation measures perform equally well; however, the simplest-looking option is presented here. It is possible to apply the same procedure with three mapping operators instead of two. Nonetheless, using three mapping operators requires additional ancilla qutrits and results in higher commutation measures due to the increased number of terms in the commutation relation. An alternative mapping approach to map away from the spin-2 subspace is to use different operators for even and odd sites. This method results in a reduced commutation measure of $\text{comm}(M)=11.5$. However, the convergence rate of the operators obtained by this approach is slow, suggesting that further conditions must be satisfied for fast convergence rates. These conditions are likely to be satisfied automatically if the operators used for odd and even sites are identical.

\section{Dephasing Noise}
\label{sec:errors}
\begin{figure*}
    \centering
    \subfloat[]{\label{fig:steering:AKLT_error:ene}\includegraphics[height=0.23\textwidth]{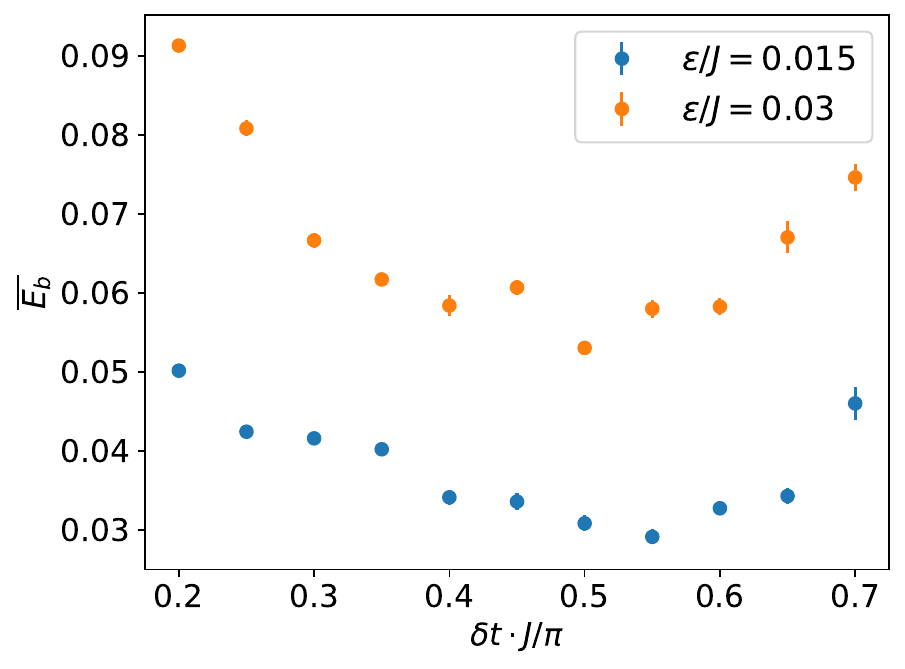}}
    \subfloat[]{\label{fig:steering:AKLT_error:errors}\includegraphics[height=0.23\textwidth]{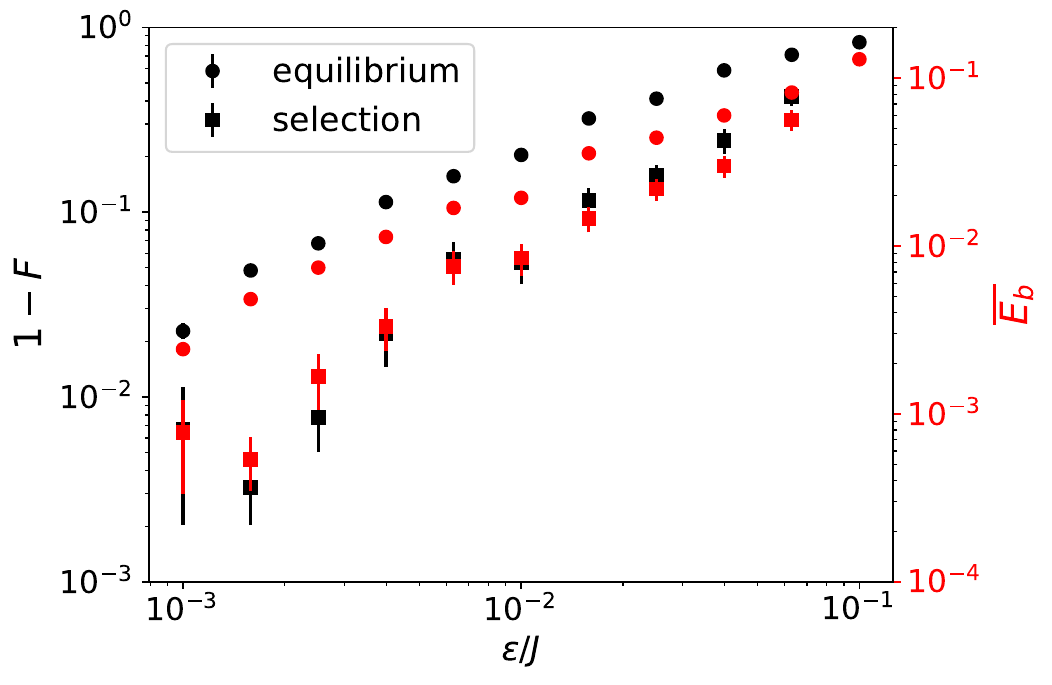}}
    \subfloat[]{\label{fig:steering:AKLT_error:tconv}\includegraphics[height=0.23\textwidth]{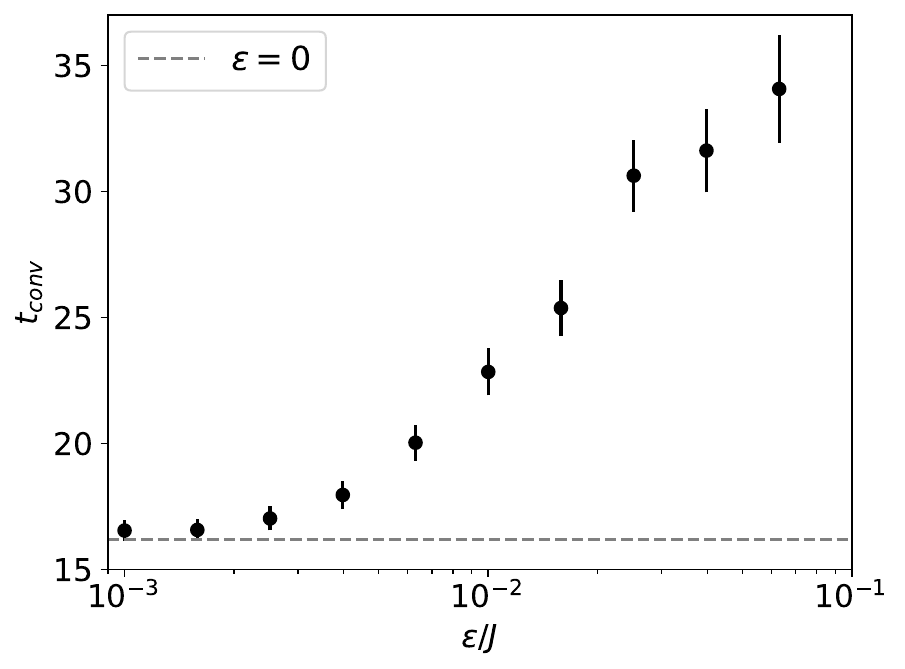}}
    \caption{The figure demonstrates the dependencies of mean bond energy $\overline{E_\text{b}}$, infidelity $1-F$, and convergence time $t_\text{conv}$ on error rates $\epsilon$ and measurement time intervals $\delta t$ for a system size of $L=10$. In subplot (a), $\epsilon$ is held constant at 0.015 and 0.03, while $\delta t$ varies. Subplots (b) and (c) employ a fixed $\delta t$ of $\frac{\pi}{2}$, with $\epsilon$ serving as the variable parameter. In (b), equilibrium values for energy per bond are calculated using $\overline{E_\text{b}}=\frac{1}{k(L-1)}\sum_{i=N-k}^N\text{Tr}(H\chi(t_i))$. The selection scheme's energies (red) and infidelities (black) are assessed by measuring observables after four consecutive evolution and measurement cycles without ancilla flips. In (c), the duration required to measure four successive cycles without any ancilla flipping is depicted for different error rates. At low error rates $t_\text{conv}$ converges to noise-free convergence time.}
    \label{fig:steering:AKLT_error}
\end{figure*} 

We test the protocol's resilience to dephasing errors, which erases quantum entanglement during the dynamics. We consider here Markovian processes described by the Lindbladian
\begin{align}
\dot{\chi} &= -i[H,\chi] +\epsilon\sum_{l,\alpha\in {\{\uparrow,0,\downarrow\}}} \left[P^\alpha_l \chi P^{\alpha\dagger}_l - \frac{1}{3}\chi\right]\, ,
\end{align}
which affects both the system and ancillas through a Lindbladian 
Here, the $P^\alpha$ operators project the system and ancillas into spin-up, spin-zero, or spin-down states. 

Dephasing is implemented by performing random projection measurements without resetting, with probability $P(\epsilon)=1-e^{-\epsilon dt}$ applied to both the system and the ancillas.

A metric for evaluating the performance of the protocol is the stationary mean energy $\overline{E_\text{b}}=\frac{1}{k(L-1)}\sum_{i=N-k}^N\text{Tr}(H\chi(t_i))$, which indicates the degree of convergence to the target state upon reaching equilibrium (see~Fig.~\ref{fig:steering:AKLT_error:ene}). For error rates of $\epsilon=0.015$ and $0.030$, the optimal $\delta t$ remains approximately $\frac{\pi}{2}$. This is consistent with the optimal $\delta t$ obtained in the absence of errors, leading to the conclusion that the optimal $\delta t$ is independent of the error rate.

It is crucial to evaluate the robustness of the protocol against different noise levels. For this purpose, both the infidelity and the mean energy per bond at equilibrium are plotted against the error rate in Fig.~\ref{fig:steering:AKLT_error:errors}. Stopping the protocol at a predetermined time when equilibrium is assumed to be reached is inefficient. Instead, the information obtained from measuring the ancillas can be used to determine when to stop the evolution of the state, this is done in a similar fashion to Matthies \etal \cite{matthies2022programmable}. The state's evolution is stopped when no ancillas have been measured in an excited state for a duration of $t_\text{wait}=4\delta t$, at which point fidelity and energy per bond are measured. Importantly, extending the waiting time would not substantially improve fidelity to the target state, as the primary factor affecting performance at large $t_\text{wait}$ is the likelihood of an error occurring during the final Hamiltonian evolution. This simple selection mechanism enhances fidelity by approximately a factor of three. This allows the self-correcting nature of the protocol to achieve even better fidelities, exceeding a fidelity of $F=0.99$ at $\epsilon=10^{-3}$, with the help of the stopping time selection scheme.

Fig.~\ref{fig:steering:AKLT_error:tconv} shows the average time for the system to exhibit no ancilla flips for $t_\text{wait}=4\delta t$. For small error rates, these values converge to the error-free convergence time discussed in Sec.~\ref{sec:convergence_time}.
For larger error rates, an exponential growth with the error rate is observed. This exponential growth is expected, as higher error rates make it exponentially less likely to encounter a period where no excited ancillas are measured.

\section{Conclusion}
In conclusion, this study investigated the potential applications of a stroboscopic map for quantum state preparation, based on the periodic resetting of ancillary degrees of freedom. We focused on the protocol introduced by Gefen \etal\cite{roy2020measurement} for unravelling quantum state steering into states with frustration-free parent Hamiltonians, and went beyond the regimes discussed there.
The AKLT state was chosen as a test case due to its simplicity, topological symmetry-protected properties, and potential usefulness for quantum computation. Although the protocol is applicable to other states with similar features, such as the cluster state, the investigation primarily centered on the AKLT state. This study provided insights into different regimes dictated by the ancilla resetting time.

Two distinct limits of the system's evolution were identified: the weak-coupling limit, resembling Markovian dynamics and described by a Lindblad equation, and the strong-coupling limit, characterized by increasing entanglement entropy production. A broad optimum between these two limits was found around $\edit{J}\delta t \simeq \frac{\pi}{2}$, a value identified in a simplified model of commuting mapping operators. In turn, a measure of non-commutativity was used to determine optimal mapping operators: the higher such a measure is, the more entanglement entropy is produced, and the slower the convergence.

In the strong-coupling limit, despite the evident peak of generated entanglement, we identified no phase transition in the model.
To this end, we introduced an entanglement-extrapolation technique and highlighted that no lattice size dependent discontinuity emerged in either the time to reach the peak or its value. We, therefore, concluded that the protocol converges also at large $\delta t$, albeit slower and slower.

Furthermore, we demonstrated that the protocol is resistant to dephasing noise, achieving a fidelity $F>0.99$ for an error rate of $\epsilon=10^{-3}$. The optimal resetting time remained approximately $\edit{J}\delta t =\frac{\pi}{2}$, and the protocol exhibited a self-correcting nature. A stopping time selection scheme based on the ancilla measurement results was introduced, which improved fidelity by approximately a factor of three.

It is worth noting that the recent work by Smith \etal\cite{smith2022deterministic} presents a method that outperforms the approach in preparing specifically the AKLT state. However, their method relies on non-local measurement-based feedback, which is not yet widely accessible on current quantum platforms. This work contributes to the understanding of alternative or complementary state preparation methods, where the roles of measurement and non-locality may be removed.

Future research directions include studying the behavior of the protocol for different Hamiltonians and target states, as well as exploring alternative strategies for choosing the mapping operators, such as incorporating feedback and using machine learning methods. Overall, the results can contribute to the development of noise-resistant state preparation routines, especially in the context of noisy intermediate-scale quantum (NISQ) devices.

\subsection*{Acknowledgements}
We thank M. Buchhold, S. Diehl, Y. Gefen, I. Gorny,  C. Koch, and K. Snizhko for inspiring discussions on the topic. We acknowledge funding by the German Federal Ministry of Education and Research (BMBF) for support under the thematic programme ``Quantum technologies -- from the basics to the market'', project number 13N16202 ``Noise in Quantum Algorithms (NiQ)''. This work was also partially funded by the Deutsche Forschungsgemeinschaft (DFG, German Research Foundation),  Project-ID 429529648 – TRR 306 QuCoLiMa (“Quantum Cooperativity of Light and Matter’’) and Project-ID 277101999, within the CRC network TRR 183. F.M. and M.R. are further supported by the DFG under Germany’s Excellence Strategy – Cluster of Excellence Matter and Light for Quantum Computing (ML4Q) EXC 2004/1 – 390534769. The authors gratefully acknowledge the Gauss Centre for Supercomputing e.V. (www.gauss-centre.eu) for funding this project by providing computing time through the John von Neumann Institute for Computing (NIC) on the GCS Supercomputer JUWELS at the J\"ulich Supercomputing Centre (JSC) (Grant NeTeNeSyQuMa) and the FZ J\"ulich for computing time on JURECA (institute project PGI-8)~\cite{JURECA2021}. 
%

\bibliographystyle{quantum}
\bibliography{literature.bib}

\begin{thebibliography}{10}

\bibitem{Preskill2018quantumcomputingin}
John Preskill.
\newblock ``Quantum {C}omputing in the {NISQ} era and beyond''.
\newblock \href{https://dx.doi.org/10.22331/q-2018-08-06-79}{{Quantum} {\bf 2},
  79}~(2018).

\bibitem{eisert2021entangling}
Jens Eisert.
\newblock ``Entangling power and quantum circuit complexity''.
\newblock Physical Review Letters {\bf 127}, 020501~(2021).
\newblock  url:~\url{https://doi.org/10.1103/physrevlett.127.020501}.

\bibitem{RevModPhys.90.015002}
Tameem Albash and Daniel~A. Lidar.
\newblock ``Adiabatic quantum computation''.
\newblock \href{https://dx.doi.org/10.1103/RevModPhys.90.015002}{Rev. Mod.
  Phys. {\bf 90}, 015002}~(2018).

\bibitem{sompet2022realizing}
Pimonpan Sompet, Sarah Hirthe, Dominik Bourgund, Thomas Chalopin, Julian Bibo,
  Joannis Koepsell, Petar Bojovi{\'c}, Ruben Verresen, Frank Pollmann,
  Guillaume Salomon, et~al.
\newblock ``Realizing the symmetry-protected haldane phase in fermi--hubbard
  ladders''.
\newblock NaturePages 1--5~(2022).
\newblock  url:~\url{https://doi.org/10.1038/s41586-022-04688-z}.

\bibitem{wei2022efficient}
Zhi-Yuan Wei, Daniel Malz, and J.~Ignacio Cirac.
\newblock ``Efficient adiabatic preparation of tensor network states''.
\newblock \href{https://dx.doi.org/10.1103/physrevresearch.5.l022037}{Physical
  Review Research{\bf 5}}~(2023).

\bibitem{PhysRevLett.95.110503}
C.~Sch\"on, E.~Solano, F.~Verstraete, J.~I. Cirac, and M.~M. Wolf.
\newblock ``Sequential generation of entangled multiqubit states''.
\newblock \href{https://dx.doi.org/10.1103/PhysRevLett.95.110503}{Phys. Rev.
  Lett. {\bf 95}, 110503}~(2005).

\bibitem{motzoi2017linear}
Felix Motzoi, Michael~P Kaicher, and Frank~K Wilhelm.
\newblock ``Linear and logarithmic time compositions of quantum many-body
  operators''.
\newblock Physical review letters {\bf 119}, 160503~(2017).
\newblock  url:~\url{https://doi.org/10.1103/physrevlett.119.160503}.

\bibitem{Poyatos:1996}
J.~F. Poyatos, J.~I. Cirac, and P.~Zoller.
\newblock ``Quantum reservoir engineering with laser cooled trapped ions''.
\newblock \href{https://dx.doi.org/10.1103/PhysRevLett.77.4728}{Phys. Rev.
  Lett. {\bf 77}, 4728--4731}~(1996).

\bibitem{Pielawa:2007}
Susanne Pielawa, Giovanna Morigi, David Vitali, and Luiz Davidovich.
\newblock ``Generation of einstein-podolsky-rosen-entangled radiation through
  an atomic reservoir''.
\newblock \href{https://dx.doi.org/10.1103/PhysRevLett.98.240401}{Phys. Rev.
  Lett. {\bf 98}, 240401}~(2007).

\bibitem{Diehl:2008}
S.~Diehl, A.~Micheli, A.~Kantian, B.~Kraus, H.~P. B{\"u}chler, and P.~Zoller.
\newblock ``Quantum states and phases in driven open quantum systems with cold
  atoms''.
\newblock \href{https://dx.doi.org/10.1038/nphys1073}{Nature Physics {\bf 4},
  878--883}~(2008).

\bibitem{Verstrate:2009}
Frank Verstraete, Michael~M. Wolf, and J.~Ignacio~Cirac.
\newblock ``Quantum computation and quantum-state engineering driven by
  dissipation''.
\newblock \href{https://dx.doi.org/10.1038/nphys1342}{Nature Physics {\bf 5},
  633--636}~(2009).

\bibitem{schirmer2010stabilizing}
SG~Schirmer and Xiaoting Wang.
\newblock ``Stabilizing open quantum systems by markovian reservoir
  engineering''.
\newblock \href{https://dx.doi.org/10.1103/physreva.81.062306}{Physical Review
  A {\bf 81}, 062306}~(2010).

\bibitem{Morigi:2015}
Giovanna Morigi, J\"urgen Eschner, Cecilia Cormick, Yiheng Lin, Dietrich
  Leibfried, and David~J. Wineland.
\newblock ``Dissipative quantum control of a spin chain''.
\newblock \href{https://dx.doi.org/10.1103/PhysRevLett.115.200502}{Phys. Rev.
  Lett. {\bf 115}, 200502}~(2015).

\bibitem{zhou2021symmetry}
Leo Zhou, Soonwon Choi, and Mikhail~D Lukin.
\newblock ``Symmetry-protected dissipative preparation of matrix product
  states''.
\newblock Physical Review A {\bf 104}, 032418~(2021).
\newblock  url:~\url{https://doi.org/10.1103/physreva.104.032418}.

\bibitem{motzoi2016backaction}
Felix Motzoi, Eli Halperin, Xiaoting Wang, K~Birgitta Whaley, and Sophie
  Schirmer.
\newblock ``Backaction-driven, robust, steady-state long-distance qubit
  entanglement over lossy channels''.
\newblock Physical Review A {\bf 94}, 032313~(2016).
\newblock  url:~\url{https://doi.org/10.1103/physreva.94.032313}.

\bibitem{smith2022deterministic}
Kevin~C. Smith, Eleanor Crane, Nathan Wiebe, and S.M. Girvin.
\newblock ``Deterministic constant-depth preparation of the aklt state on a
  quantum processor using fusion measurements''.
\newblock \href{https://dx.doi.org/10.1103/prxquantum.4.020315}{PRX Quantum{\bf
  4}}~(2023).

\bibitem{tantivasadakarn2021long}
Nathanan Tantivasadakarn, Ryan Thorngren, Ashvin Vishwanath, and Ruben
  Verresen.
\newblock ``Long-range entanglement from measuring symmetry-protected
  topological phases''~(2021).
\newblock  url:~\url{https://arxiv.org/abs/2112.01519}.

\bibitem{sayrin2011real}
Cl{\'e}ment Sayrin, Igor Dotsenko, Xingxing Zhou, Bruno Peaudecerf, Th{\'e}o
  Rybarczyk, S{\'e}bastien Gleyzes, Pierre Rouchon, Mazyar Mirrahimi, Hadis
  Amini, Michel Brune, et~al.
\newblock ``Real-time quantum feedback prepares and stabilizes photon number
  states''.
\newblock Nature {\bf 477}, 73--77~(2011).
\newblock  url:~\url{https://doi.org/10.1038/nature10376}.

\bibitem{vijay2012stabilizing}
R~Vijay, Chris Macklin, DH~Slichter, SJ~Weber, KW~Murch, Ravi Naik, Alexander~N
  Korotkov, and Irfan Siddiqi.
\newblock ``Stabilizing rabi oscillations in a superconducting qubit using
  quantum feedback''.
\newblock Nature {\bf 490}, 77--80~(2012).
\newblock  url:~\url{https://doi.org/10.1038/nature11505}.

\bibitem{riste2013deterministic}
D~Riste, M~Dukalski, CA~Watson, G~De~Lange, MJ~Tiggelman, Ya~M Blanter,
  Konrad~W Lehnert, RN~Schouten, and L~DiCarlo.
\newblock ``Deterministic entanglement of superconducting qubits by parity
  measurement and feedback''.
\newblock Nature {\bf 502}, 350--354~(2013).
\newblock  url:~\url{https://doi.org/10.1038/nature12513}.

\bibitem{mabuchi2009continuous}
Hideo Mabuchi.
\newblock ``Continuous quantum error correction as classical hybrid control''.
\newblock New Journal of Physics {\bf 11}, 105044~(2009).
\newblock  url:~\url{https://doi.org/10.1088/1367-2630/11/10/105044}.

\bibitem{kerckhoff2010designing}
Joseph Kerckhoff, Hendra~I Nurdin, Dmitri~S Pavlichin, and Hideo Mabuchi.
\newblock ``Designing quantum memories with embedded control: photonic circuits
  for autonomous quantum error correction''.
\newblock Physical Review Letters {\bf 105}, 040502~(2010).
\newblock  url:~\url{https://doi.org/10.1103/physrevlett.105.040502}.

\bibitem{martin2015deterministic}
Leigh Martin, Felix Motzoi, Hanhan Li, Mohan Sarovar, and K~Birgitta Whaley.
\newblock ``Deterministic generation of remote entanglement with active quantum
  feedback''.
\newblock Physical Review A {\bf 92}, 062321~(2015).
\newblock  url:~\url{https://doi.org/10.1103/physreva.92.062321}.

\bibitem{google2023suppressing}
Google~Quantum AI.
\newblock ``Suppressing quantum errors by scaling a surface code logical
  qubit''.
\newblock Nature {\bf 614}, 676--681~(2023).
\newblock  url:~\url{https://www.nature.com/articles/s41586-022-05434-1}.

\bibitem{BurgarthPRA2007}
Daniel Burgarth and Vittorio Giovannetti.
\newblock ``Mediated homogenization''.
\newblock \href{https://dx.doi.org/10.1103/PhysRevA.76.062307}{Phys. Rev. A
  {\bf 76}, 062307}~(2007).

\bibitem{BurgarthPRL2007}
Daniel Burgarth and Vittorio Giovannetti.
\newblock ``Full control by locally induced relaxation''.
\newblock \href{https://dx.doi.org/10.1103/PhysRevLett.99.100501}{Phys. Rev.
  Lett. {\bf 99}, 100501}~(2007).

\bibitem{matthies2022programmable}
Anne Matthies, Mark Rudner, Achim Rosch, and Erez Berg.
\newblock ``Programmable adiabatic demagnetization for systems with trivial and
  topological excitations''~(2022).
\newblock  url:~\url{https://arxiv.org/abs/2210.17256}.

\bibitem{roy2020measurement}
Sthitadhi Roy, JT~Chalker, IV~Gornyi, and Yuval Gefen.
\newblock ``Measurement-induced steering of quantum systems''.
\newblock Physical Review Research {\bf 2}, 033347~(2020).
\newblock  url:~\url{https://doi.org/10.1103/physrevresearch.2.033347}.

\bibitem{moore2001parallel}
Cristopher Moore and Martin Nilsson.
\newblock ``Parallel quantum computation and quantum codes''.
\newblock SIAM journal on computing {\bf 31}, 799--815~(2001).
\newblock  url:~\url{https://doi.org/10.1137/s0097539799355053}.

\bibitem{van2005fast}
Rodney Van~Meter and Kohei~M Itoh.
\newblock ``Fast quantum modular exponentiation''.
\newblock Physical Review A {\bf 71}, 052320~(2005).
\newblock  url:~\url{https://doi.org/10.1103/physreva.71.052320}.

\bibitem{draper2004logarithmic}
Bhaskar Gaur, Edgard Mu{\~n}oz-Coreas, and Himanshu Thapliyal.
\newblock ``A logarithmic depth quantum carry-lookahead modulo (2n - 1)
  adder''.
\newblock In Proceedings of the Great Lakes Symposium on VLSI 2023.
\newblock \href{https://dx.doi.org/10.1145/3583781.3590205}{Pages 125--130}.
\newblock ~(2023).

\bibitem{jacobs2014coherent}
Kurt Jacobs, Xiaoting Wang, and Howard~M Wiseman.
\newblock ``Coherent feedback that beats all measurement-based feedback
  protocols''.
\newblock \href{https://dx.doi.org/10.1088/1367-2630/16/7/073036}{New Journal
  of Physics {\bf 16}, 073036}~(2014).

\bibitem{rivas2010entanglement}
{\'A}ngel Rivas, Susana~F Huelga, and Martin~B Plenio.
\newblock ``Entanglement and non-markovianity of quantum evolutions''.
\newblock Physical review letters {\bf 105}, 050403~(2010).
\newblock  url:~\url{https://doi.org/10.1103/physrevlett.105.050403}.

\bibitem{verresen2017one}
Ruben Verresen, Roderich Moessner, and Frank Pollmann.
\newblock ``One-dimensional symmetry protected topological phases and their
  transitions''.
\newblock Physical Review B {\bf 96}, 165124~(2017).
\newblock  url:~\url{https://doi.org/10.1103/physrevb.96.165124}.

\bibitem{pollmann2012detection}
Frank Pollmann and Ari~M Turner.
\newblock ``Detection of symmetry-protected topological phases in one
  dimension''.
\newblock Physical review b {\bf 86}, 125441~(2012).
\newblock  url:~\url{https://doi.org/10.1103/physrevb.86.125441}.

\bibitem{brennen2008measurement}
Gavin~K Brennen and Akimasa Miyake.
\newblock ``Measurement-based quantum computer in the gapped ground state of a
  two-body hamiltonian''.
\newblock Physical review letters {\bf 101}, 010502~(2008).
\newblock  url:~\url{https://doi.org/10.1103/physrevlett.101.010502}.

\bibitem{Filipowicz:1986}
P.~Filipowicz, J.~Javanainen, and P.~Meystre.
\newblock ``Theory of a microscopic maser''.
\newblock \href{https://dx.doi.org/10.1103/PhysRevA.34.3077}{Phys. Rev. A {\bf
  34}, 3077--3087}~(1986).

\bibitem{Slosser:1990}
John~J. Slosser and Pierre Meystre.
\newblock ``Tangent and cotangent states of the electromagnetic field''.
\newblock \href{https://dx.doi.org/10.1103/PhysRevA.41.3867}{Phys. Rev. A {\bf
  41}, 3867--3874}~(1990).

\bibitem{Briegel:1995}
Hans-J\"urgen Briegel and Berthold-Georg Englert.
\newblock ``Macroscopic dynamics of a maser with non-poissonian injection
  statistics''.
\newblock \href{https://dx.doi.org/10.1103/PhysRevA.52.2361}{Phys. Rev. A {\bf
  52}, 2361--2375}~(1995).

\bibitem{Wellens:2000}
Thomas Wellens, Andreas Buchleitner, Burkhard K\"ummerer, and Hans Maassen.
\newblock ``Quantum state preparation via asymptotic completeness''.
\newblock \href{https://dx.doi.org/10.1103/PhysRevLett.85.3361}{Phys. Rev.
  Lett. {\bf 85}, 3361--3364}~(2000).

\bibitem{Pielawa:2010}
Susanne Pielawa, Luiz Davidovich, David Vitali, and Giovanna Morigi.
\newblock ``Engineering atomic quantum reservoirs for photons''.
\newblock \href{https://dx.doi.org/10.1103/PhysRevA.81.043802}{Phys. Rev. A
  {\bf 81}, 043802}~(2010).

\bibitem{Hartmann:2017}
M~Hartmann, D~Poletti, M~Ivanchenko, S~Denisov, and P~H\"anggi.
\newblock ``Asymptotic floquet states of open quantum systems: the role of
  interaction''.
\newblock \href{https://dx.doi.org/10.1088/1367-2630/aa7ceb}{New Journal of
  Physics {\bf 19}, 083011}~(2017).

\bibitem{Weidinger:1999}
M.~Weidinger, B.~T.~H. Varcoe, R.~Heerlein, and H.~Walther.
\newblock ``Trapping states in the micromaser''.
\newblock \href{https://dx.doi.org/10.1103/PhysRevLett.82.3795}{Phys. Rev.
  Lett. {\bf 82}, 3795--3798}~(1999).

\bibitem{Varcoe:2000}
B.~T.~H. Varcoe, S.~Brattke, M.~Weidinger, and H.~Walther.
\newblock ``Preparing pure photon number states of the radiation field''.
\newblock \href{https://dx.doi.org/10.1038/35001526}{Nature {\bf 403},
  743--746}~(2000).

\bibitem{Morigi:1997}
G.~Morigi, J.~I. Cirac, M.~Lewenstein, and P.~Zoller.
\newblock ``Ground-state laser cooling beyond the lamb-dicke limit''.
\newblock \href{https://dx.doi.org/10.1209/epl/i1997-00306-3}{Europhysics
  Letters {\bf 39}, 13}~(1997).

\bibitem{Morigi:1998}
G.~Morigi, J.~I. Cirac, K.~Ellinger, and P.~Zoller.
\newblock ``Laser cooling of trapped atoms to the ground state: A dark state in
  position space''.
\newblock \href{https://dx.doi.org/10.1103/PhysRevA.57.2909}{Phys. Rev. A {\bf
  57}, 2909--2914}~(1998).

\bibitem{Dalibard:1992}
Jean Dalibard, Yvan Castin, and Klaus M\o{}lmer.
\newblock ``Wave-function approach to dissipative processes in quantum
  optics''.
\newblock \href{https://dx.doi.org/10.1103/PhysRevLett.68.580}{Phys. Rev. Lett.
  {\bf 68}, 580--583}~(1992).

\bibitem{Dum:1992}
R.~Dum, P.~Zoller, and H.~Ritsch.
\newblock ``Monte carlo simulation of the atomic master equation for
  spontaneous emission''.
\newblock \href{https://dx.doi.org/10.1103/PhysRevA.45.4879}{Phys. Rev. A {\bf
  45}, 4879--4887}~(1992).

\bibitem{Cubitt:2003}
T.~S. Cubitt, F.~Verstraete, W.~D\"ur, and J.~I. Cirac.
\newblock ``Separable states can be used to distribute entanglement''.
\newblock \href{https://dx.doi.org/10.1103/PhysRevLett.91.037902}{Phys. Rev.
  Lett. {\bf 91}, 037902}~(2003).

\bibitem{Roldan:2017}
\'Edgar Rold\'an and Shamik Gupta.
\newblock ``Path-integral formalism for stochastic resetting: Exactly solved
  examples and shortcuts to confinement''.
\newblock \href{https://dx.doi.org/10.1103/PhysRevE.96.022130}{Phys. Rev. E
  {\bf 96}, 022130}~(2017).

\bibitem{Mukherjee:2018}
B.~Mukherjee, K.~Sengupta, and Satya~N. Majumdar.
\newblock ``Quantum dynamics with stochastic reset''.
\newblock \href{https://dx.doi.org/10.1103/PhysRevB.98.104309}{Phys. Rev. B
  {\bf 98}, 104309}~(2018).

\bibitem{Yin:2023}
R.~Yin and E.~Barkai.
\newblock ``Restart expedites quantum walk hitting times''.
\newblock \href{https://dx.doi.org/10.1103/PhysRevLett.130.050802}{Phys. Rev.
  Lett. {\bf 130}, 050802}~(2023).

\bibitem{haegeman2011time}
Jutho Haegeman, J~Ignacio Cirac, Tobias~J Osborne, Iztok Pi{\v{z}}orn, Henri
  Verschelde, and Frank Verstraete.
\newblock ``Time-dependent variational principle for quantum lattices''.
\newblock Physical review letters {\bf 107}, 070601~(2011).
\newblock  url:~\url{https://doi.org/10.1007/3-540-10579-4_20}.

\bibitem{daley2014quantum}
Andrew~J. Daley.
\newblock ``Quantum trajectories and open many-body quantum systems''.
\newblock \href{https://dx.doi.org/10.1080/00018732.2014.933502}{Advances in
  Physics {\bf 63}, 77–149}~(2014).

\bibitem{JURECA2021}
J{\"u}lich~Supercomputing Centre.
\newblock ``Jureca: Data centric and booster modules implementing the modular
  supercomputing architecture at j{\"u}lich supercomputing centre''.
\newblock \href{https://dx.doi.org/10.17815/jlsrf-7-182}{Journal of large-scale
  research facilities {\bf 7}, A182}~(2021).

\bibitem{garcia2013spectral}
Artur Garcia-Saez, Valentin Murg, and Tzu-Chieh Wei.
\newblock ``Spectral gaps of affleck-kennedy-lieb-tasaki hamiltonians using
  tensor network methods''.
\newblock Physical Review B {\bf 88}, 245118~(2013).
\newblock  url:~\url{https://doi.org/10.1103/physrevb.88.245118}.

\end{thebibliography}
\onecolumn
\appendix

\section{Imaginary time evolution}
\label{appendix:imaginary_time_evolution}
\begin{figure}[ht]
    \centering
    \includegraphics[width=0.45\textwidth]{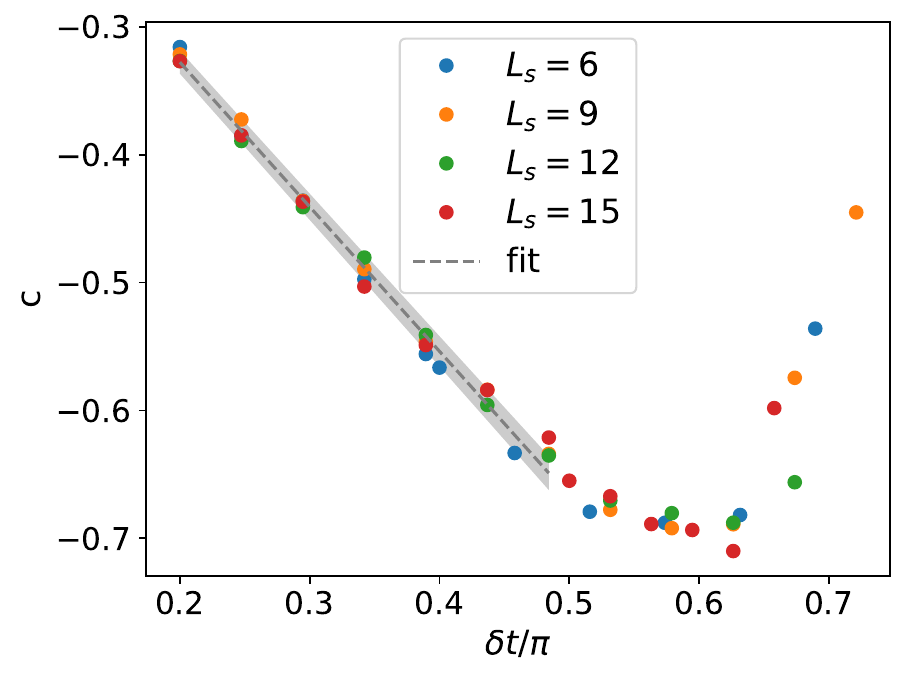}
    \caption{The figure shows the exponential coefficient $c$ of the energy evolution $E(t)=e^{ct+a}$ for different values of $\delta t$ after the system has reached the convergence time $t_\text{conv}$. A linear fit of $c(\delta t)$ for $L_s=15$ gives a slope of $\widetilde{\Delta E}=-0.351 \pm 0.009$, which is consistent with the theoretical value for the energy gap.}
    \label{fig:steering:imaginary_time_evolution}
\end{figure}
This section presents an analysis of the dynamics' evolution beyond $t_\text{conv}$, which represents the time when the ancillas no longer flip into an excited state. The lack of ancilla flipping effectively causes only the imaginary time evolution term in Eq.~\ref{eq:steering:lindblad_limit:ham} to act on the state. If this assumption holds true, the energy gap of the AKLT Hamiltonian can be readily obtained. By comparing this energy gap to values reported in the literature, we can confirm the accuracy of our assumption regarding the protocol reducing to imaginary time evolution.

At $t_\text{conv}$, the state closely approximates the AKLT state, allowing us to assume a low-temperature thermal state:
\begin{align}
\rho=\frac{\ket{E_0}\bra{E_0}}{{1 + e^{-\Delta E \beta_0}}} + \ket{E_1}\bra{E_1} \frac{e^{-\Delta E \beta_0}}{1 + e^{-\Delta E \beta_0}}
\end{align}
where $\ket{E_1}$ denotes the first excited state, $\Delta E$ represents the energy gap, and $\beta_0$ is the inverse temperature.
Evolving the state with the imaginary time evolution term in Eq.~\ref{eq:steering:lindblad_limit:ham} results in:
\begin{align}
e^{-\frac{t}{2} H_\text{AKLT} \delta t}\rho e^{-\frac{t}{2} H_\text{AKLT} \delta t}&= \frac{\ket{E_0}\bra{E_0}}{1 + e^{-\Delta E \beta_0}}+\frac{\ket{E_1}\bra{E_1}}{1 + e^{\Delta E (\beta_0 + \delta t \cdot t)}} \\
E(t) &= \frac{\Delta E}{1 + e^{\Delta E (\beta_0 + \delta t \cdot t)}}
\end{align}
Consequently, if $\beta_0$ is sufficiently large, the energy should scale as $E(t)\propto e^{-\Delta E \delta t \cdot t}$.
To compare this relationship with the simulation data, the exponential factor $c(\delta t)$ can be calculated by fitting the exponential $E(t)=e^{ct+a}$ to the energy of the trajectories after $t_\text{conv}$. This fitting produces the data shown in Fig.~\ref{fig:steering:imaginary_time_evolution}, which displays a linear decrease until it starts to flatten near $\delta t = \frac{\pi}{2}$. The linear part was fit with $c(\delta t)\approx -\delta t \widetilde{\Delta E} + b$, yielding $\widetilde{\Delta E}=0.351 \pm 0.009$. This value is consistent with the energy gap of the AKLT Hamiltonian, $\Delta E = 0.3501$ \cite{garcia2013spectral}, thus corroborating our comprehension of the dynamics.

\section{Entanglement extrapolation}
\label{appendix:entanglement_extrapolation}
\begin{figure}
    \centering
    \subfloat[]{\label{fig:steering:entanglement_extraploation:example}\includegraphics[width=0.33\textwidth]{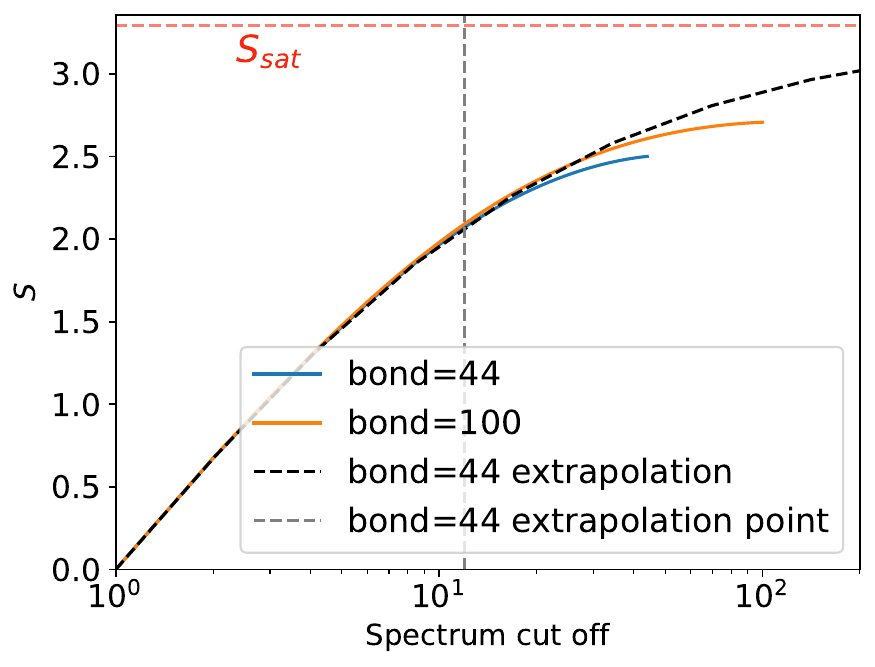}}
    \subfloat[]{\label{fig:steering:entanglement_extraploation:example_44}\includegraphics[width=0.33\textwidth]{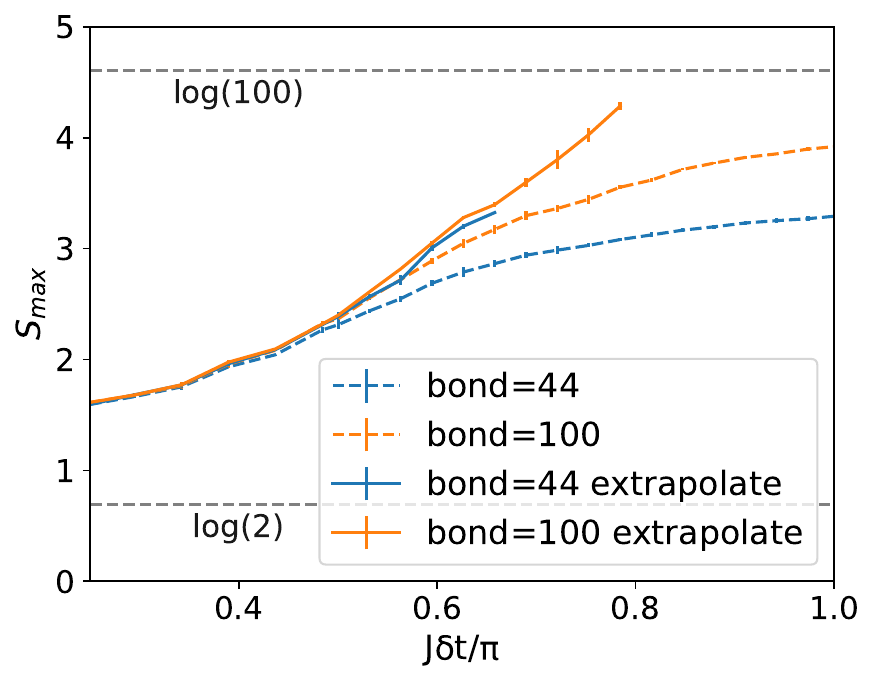}}
    \subfloat[]{\label{fig:steering:entanglement_extraploation:example_lat}\includegraphics[width=0.33\textwidth]{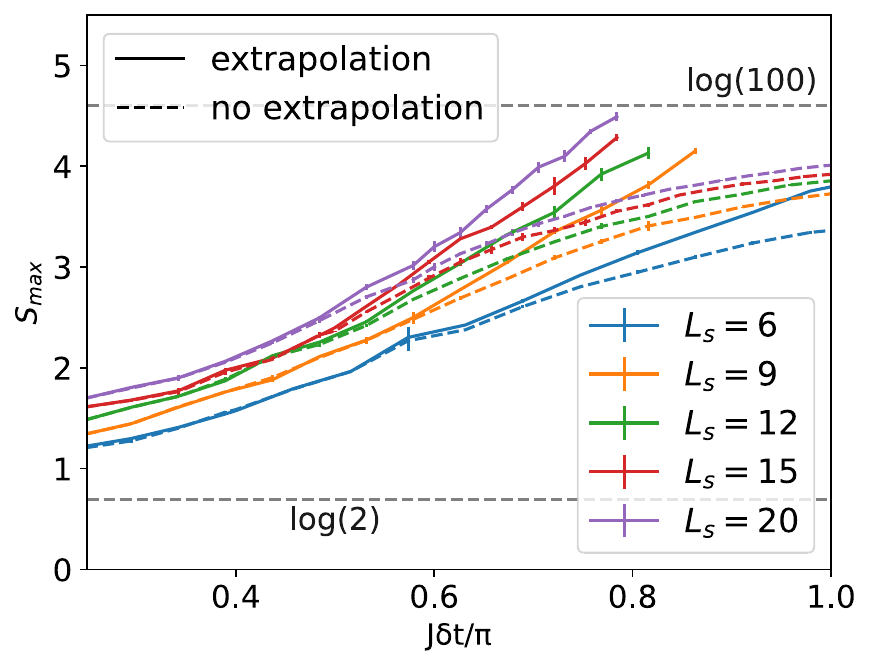}}
    \caption{(a)  The entanglement is extrapolated from a bond dimension of 44 using Eq.~\ref{eq:steering:extrapolation}. (b) The maximal entanglement achieved in a simulation for a system size of $L_s=15$ is shown. By comparing extrapolated and non-extrapolated results, we can see that the extrapolated values at a bond dimension of 44 match the more expensive simulation at a bond dimension of 100. (c) The same plot as in (b) is shown, but this time comparing different system sizes $L_s$.}
    \label{fig:steering:entanglement_extraploation}
\end{figure}

In this section, we introduce a novel technique to extrapolate the entanglement of a Matrix Product State (MPS). This method is crucial for determining the validity duration of an MPS simulation and for addressing the question of whether an entanglement phase transition can be observed.

The entanglement of a quantum system can be computed by bisecting the system and performing a Singular Value Decomposition (SVD) $\ket{\psi}=\sum_i^{2^{L/2}} \lambda_i \ket{\psi^i_L} \otimes \ket{\psi^i_R}$. Once the singular values are obtained, the entanglement entropy can be calculated using the expression $S=\sum_i^{2^{L/2}} \lambda_i^2 \log(\lambda_i^2)$. The tensor network technique involves retaining only the largest $\lambda_i$, so that the new entanglement is
\begin{align}
p_i =&\frac{\lambda_i^2}{\sum_i^{\text{cut}} \lambda_i^2} \nonumber \\
S(\text{cut})=&\sum_i^{\text{cut}} p_i \log(p_i).
\end{align}
If the discarded $\lambda_i$ are excessively large, the simulation is no longer valid.

Our approach approximates the magnitude of the discarded eigenvalues through a four-step process. First, simulations are conducted at both small and large bond dimensions. Second, the entanglement is computed for various cutting points at both bond dimensions. Third, the saturation entanglement entropy is extrapolated by fitting $S(\text{cut})$ with $\text{cut}<0.9 \text{bond}_\text{small}$:
\begin{align}
S(\text{cut})=S_\text{sat} \tanh\biggl(
\frac{\log(\text{cut}) \sigma_1}{S_\text{sat}}
&+\frac{\log(\text{cut})^2 \sigma_2}{S_\text{sat}} \nonumber \\
&+\frac{\sqrt{\log(\text{cut})} \sigma_3}{S_\text{sat}}
\biggr). \label{eq:steering:extrapolation}
\end{align}
Lastly, the resulting saturation points $S_\text{sat}$ are compared for both bond dimensions. If they match, the extrapolation is deemed valid, and a new extrapolation is performed with $S(\text{cut})$ with $\text{cut}<0.9 \text{bond}_\text{large}$.

It is important to note that the magnitude of the $\sigma_1$ term indicates the number of eigenvalues needed to reach the saturation point, while the terms $\sigma_2$ and $\sigma_3$ slightly adjust the curvature of the saturation curve and are nearly zero for most extrapolations. Adding more terms to the expansion does not affect the saturation point $S_\text{sat}$. An example of this procedure is illustrated in Fig.~\ref{fig:steering:entanglement_extraploation:example}.

In Fig.~\ref{fig:steering:entanglement_extraploation:example_44}, the maximal entanglement achieved in the simulation $S_\text{max}$ is presented for bond dimensions of 44 and 100, with extrapolation both enabled and disabled. It is evident that although the non-extrapolated values diverge, the extrapolated ones yield consistent results. Given that conducting a simulation at bond dimension 100 is eight times more resource-intensive than at bond dimension 44, these findings are promising.

Lastly, Fig.~\ref{fig:steering:entanglement_extraploation:example_lat} displays the maximal entropy observed in the simulation for various reset intervals $\delta t$ across different lattice sizes. The values of $S_\text{max}$ without employing the extrapolation technique begin to flatten as they approach the maximal entropy that the MPS can represent $\log(100)$. In contrast, the extrapolated values increase linearly until their validity can no longer be verified. This demonstrates the importance of applying the extrapolation technique to ensure the reliability of the results.

\section{Different initial states}
\begin{figure*}
    \centering
    \subfloat[]{\label{fig:steering:different_init:energy}\includegraphics[width=0.45\textwidth]{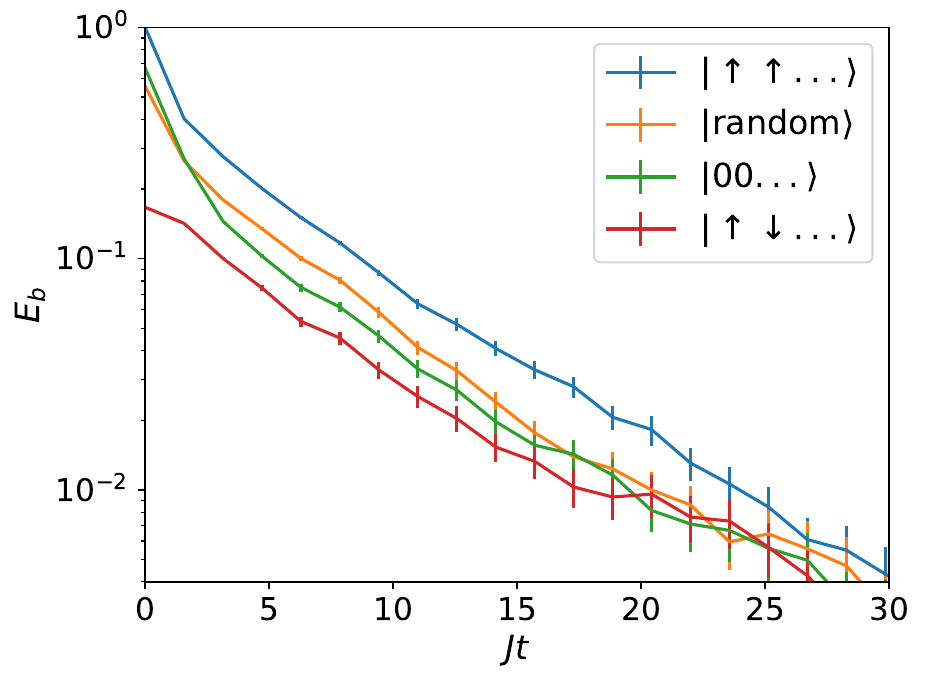}}
    \subfloat[]{\label{fig:steering:different_init:entropy}\includegraphics[width=0.45\textwidth]{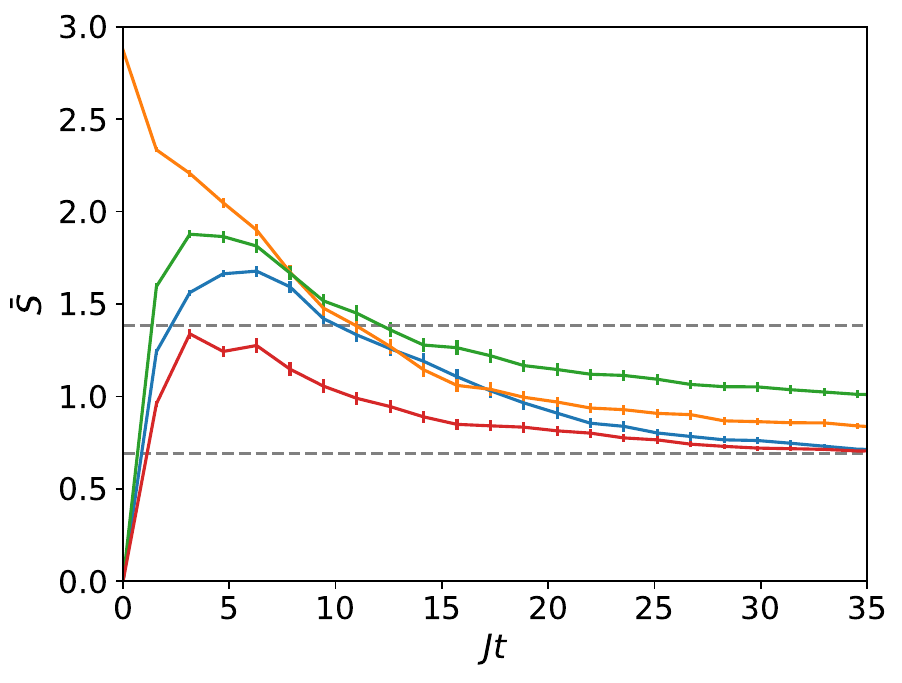}}
    
    \caption{The figure shows the measurement-induced steering of three different product states (blue: all spins pointing upwards, green: all spins in the zero component, and red: alternating spin up and down), as well as a random initial MPS state with a bond dimension of 44 (yellow). Panel (a) shows that states with lower initial energy converge faster, as indicated by the smoothed energy per bond $E_b$. Panel (b) reveals that the average bipartite entanglement entropy increases and then decreases for the product states but steadily decreases for the random states, resulting in entropies between $\log(2)$ and $\log(4)$. This indicates a different mixture of the four AKLT states based on the initial state.}
    \label{fig:steering:different_init}
\end{figure*}
In this appendix, we explore the robustness of our results by considering four different initial states. These states consist of:  all spins pointing up (blue); all spins in the zero component (green); alternating spin up and down (red); and a random initial MPS state with a bond dimension of 44 (yellow). As seen in Fig. 8(a), states with lower initial energy converge faster, as expected since they are closer to the ground state. Fig. 8(b) shows that the average bipartite entanglement entropy increases and then decreases for the product states, but steadily decreases for the random states. This results in entropies between $\log(2)$ and $\log(4)$, indicating a different mixture of the four AKLT states depending on the initial state. We attribute this to the fact that the product state with all spin components in the zero state is closer to the AKLT state in which the two edge modes are entangled.
\end{document}